\documentclass[journal]{IEEEtran}

\usepackage{cite}
\usepackage{amsmath,amssymb,amsfonts}
\usepackage{amsthm}
\usepackage{graphicx}
\usepackage{textcomp}
\usepackage{xcolor}
\usepackage{booktabs}
\usepackage{multirow}
\usepackage{tabularx}
\usepackage{comment}
\usepackage{hyperref}
\usepackage{subcaption} 
\usepackage{tikz}
\usetikzlibrary{positioning,arrows.meta,fit,backgrounds,calc}
\usepackage{graphicx}
\graphicspath{{./Fig/}} 

\newcounter{mycounter}

\begin{document}

\title{ARGUS: Agentic Reasoning over Grounded, Unified Semantics for Autonomous Vehicle Networks}

\author{Shavbo~Salehi,
        Pedro~Enrique~Iturria-Rivera,
        Medhat~Elsayed,
        Majid~Bavand,
        Yigit~Ozcan,
        and~Melike~Erol-Kantarci,~\IEEEmembership{Fellow, IEEE}
\thanks{Shavbo Salehi and Melike Erol-Kantarci are with the School of Electrical Engineering and Computer Science, University of Ottawa, Ottawa, ON K1N 6N5, Canada (e-mail: ssale038@uottawa.ca; melike.erolkantarci@uottawa.ca).}
\thanks{Pedro Enrique Iturria-Rivera, Medhat Elsayed, Majid Bavand, and Yigit Ozcan are with Ericsson, Ottawa, K2K 2V6, Canada(e-mail:pedro.iturria.rivera;medhat.elsayed@ericsson.com; majid.bavand@ericsson.com; yigit.ozcan@ericsson.com) 
}}

\maketitle

\begin{abstract}
Semantic communication marks a fundamental evolution for future artificial intelligence (AI)-native wireless networks, replacing bit-level reconstruction with task-oriented information exchange. In vehicular networks, where high mobility and strict latency make raw data exchange prohibitive, transmitting semantic meaning is vital for cooperative driving and safety. However, current systems rarely extend beyond basic semantic encoding and transmission, lacking the autonomous reasoning, multi-agent coordination, and adaptive decision-making needed to support these complex vehicle interactions. In this paper, we propose ARGUS (Agentic Reasoning over Grounded, Unified Semantics), an agentic semantic communication framework for autonomous vehicle networks that integrates large language-based models (LLMs) with neuro-symbolic causal reasoning and adaptive semantic communication into a unified multi-agent architecture. Each vehicle employs a vision-language model (VLM) to transform multimodal sensory observations into compact semantic representations, while an edge-based LLM agent performs semantic reasoning, correlates information across vehicles, and orchestrates cooperative decision-making. To further enhance network efficiency and reliability, the framework incorporates novelty-driven semantic adaptation for bandwidth-aware communication and a neuro-symbolic causal reasoning module for predictive hazard awareness. Compared across diverse baselines, spanning raw sensor transmission, rule-based semantic filtering, and VLM-based methods, ARGUS significantly reduces communication overhead while enhancing perception accuracy. ARGUS reduces average uplink bandwidth by 86\%, while raising the hazard-detection rate from 0.69 to 0.88, and $6\times$ higher detection throughput and the lowest end-to-end (E2E) latency. These findings demonstrate the viability of agentic AI for bandwidth-constrained, mission-critical vehicular communications. 

\end{abstract}

\begin{IEEEkeywords}
Agentic AI, LLMs, Neuro-Symbolic Reasoning, Semantic Communication, Vehicular Networks, VLMs.
\end{IEEEkeywords}

\section{Introduction}

Future 6G networks are expected to support large-scale deployments of connected and autonomous vehicles that continuously exchange information to enable cooperative perception, hazard awareness, and autonomous decision-making \cite{saad2019vision,strinati20196g}. Modern autonomous vehicles generate massive amounts of multimodal sensor data from cameras, LiDARs, radars, and onboard perception systems. Conventional Vehicle-to-Everything (V2X) communication architectures transmit raw or compressed sensor data between vehicles and roadside infrastructure to support collaborative driving applications \cite{wan2025systematic}. However, as the number of connected vehicles increases, communication requirements rapidly exceed available wireless resources. Consequently, transmitting sensor observations at the bit level becomes increasingly inefficient and limits the scalability of future intelligent transportation systems \cite{hou2025enhancing}.

Semantic communication has emerged as a broader paradigm for beyond-5G and 6G wireless networks, aiming to improve communication efficiency by transmitting task-relevant information rather than reproducing all source data
\cite{salehi2025genai}. This paradigm is particularly relevant to bandwidth-constrained V2X networks, where reducing redundant transmissions can help alleviate communication overhead while preserving information relevant to cooperative driving tasks. Unlike conventional communication systems that aim to reconstruct transmitted bits with high fidelity, semantic communication focuses on delivering the relevant meaning to a given task and omitting redundant details. Using shared knowledge and contextual understanding, receivers can reconstruct situational awareness from smaller communication payloads. Recent advances in large language models (LLMs) and generative artificial intelligence (AI) have enabled semantic communication frameworks for text, image, video, speech, and multimodal data transmission \cite{salehi2025llm,salehi2025genai,tian2026large}. These approaches have demonstrated substantial gains in communication efficiency and robustness under wireless impairments. Nevertheless, most existing semantic communication systems remain limited to encoder and decoder architectures that focus primarily on semantic compression and reconstruction, with limited capabilities for autonomous reasoning, contextual adaptation, and cooperative decision-making \cite{yang2022semantic}.

Recent breakthroughs in foundation models have enabled a new generation of intelligent agentic systems capable of perception, reasoning, planning, memory management, and autonomous decision-making \cite{tian2025large}. In particular, vision-language models (VLMs) provide powerful multimodal scene understanding capabilities by jointly processing visual and textual information, while LLMs enable high-level reasoning and coordination across distributed entities \cite{liu2024improved}. These capabilities create new opportunities for semantic communication systems in which communicating agents can actively determine what information should be transmitted, when it should be transmitted, and how communication resources should be allocated according to application objectives \cite{salehi2025genai}. Despite progress in semantic communication, multimodal foundation models, and LLM-based autonomous agents, their unified integration into semantic communication frameworks for cooperative autonomous vehicle networks has received limited attention in existing literature \cite{yao2022react,qin2024toolllm}.

Existing semantic communication frameworks also generally process observations independently at each transmission interval. However, in highly structured driving environments, many scene elements evolve gradually and remain predictable over time. Therefore, repeatedly transmitting of complete semantic descriptions introduces substantial redundancy and unnecessary communication overhead \cite{getu2025semantic}. Recent advances in neuro-symbolic AI and causal reasoning provide mechanisms for representing dynamic environments through structured causal relationships rather than independent observations \cite{deng2025deep}. By communicating only changes in the underlying causal structure of a scene, semantic communication systems can reduce transmission requirements while preserving information relevant to situational awareness. 

However, reducing the size of individual semantic messages alone does not fully address the communication and coordination challenges of cooperative autonomous driving. Vehicles may observe the same scene elements from different viewpoints, while only a subset of observations may contain new or safety-critical information. Furthermore, in dense vehicular environments, independently compressing observations on each vehicle leaves two critical issues unaddressed: first, an isolated vehicle cannot verify whether its local observation is corroborated by neighboring vehicles viewing the same event from different perspectives; second, no vehicle has global visibility into the shared uplink channel to coordinate bandwidth contention across the fleet. Consequently, purely local decision-making either over-allocates bandwidth to benign yet visually complex scenes, or drops critical reports under severe interference. This motivates semantic communication architectures that jointly integrate perception, reasoning, prediction, and adaptive wireless transmission.
Motivated by these observations, this paper proposes ARGUS, an agentic semantic communication framework for autonomous vehicle networks operating over bandwidth-constrained V2X links. In ARGUS's architecture, each vehicle employs a VLM-based agent to transform multimodal sensor observations into compact semantic scene descriptions. These semantic messages are transmitted to a roadside unit (RSU), where an LLM-based orchestration agent performs semantic reasoning, correlates cross-vehicle observations, and dynamically schedules network resources. Rather than targeting sub-millisecond emergency maneuvers, ARGUS focuses on macro-level cooperative monitoring, such as detecting emerging bottlenecks, lane obstructions, and temporary work zones, where edge-level reasoning latencies are readily acceptable. To balance edge intelligence with bandwidth constraints, the framework integrates three mechanisms; $i$) novelty-aware transmission adaptation, 
which adjusts vehicle reporting rates based on temporal scene divergence to suppress static traffic redundancy; $ii$) neuro-symbolic causal reasoning,
which grounds semantic descriptions into structured driving concepts to enable traceable hazard inference and cross-vehicle corroboration; and $iii$) diffusion-based semantic encoding
, which ensures graceful message degradation under adverse wireless channel fading. The main contributions of this paper are summarized as follows:

\begin{itemize}

\item \textbf{An agentic semantic communication architecture:} We propose ARGUS, which integrates distributed VLM-based vehicle agents with an LLM-based orchestration agent at the RSU. Vehicle agents generate compact semantic representations, while the RSU performs cross-vehicle reasoning, observation correlation, and network-level resource scheduling for cooperative situational awareness. 

\item \textbf{An adaptive semantic scene communication mechanism:} We develop a framework that replaces raw sensor transmissions with compact semantic scene descriptions and incorporates novelty-driven adaptation to dynamically adjust communication rates according to scene changes and environmental dynamics.

\item \textbf{A neuro-symbolic causal reasoning framework for reducing redundancy:} ARGUS employs causal scene graphs to represent traffic entities and their relationships, enabling graph-difference updates that communicate only novel and causally relevant information while minimizing redundant transmissions.

\item \textbf{A robust semantic transmission framework and large-scale evaluation:} We design a diffusion-based joint source-channel coding mechanism for reliable semantic transmission under wireless impairments and demonstrate substantial bandwidth savings while maintaining hazard awareness and situational understanding in large-scale autonomous vehicle scenarios.
\end{itemize}

\section{Literature Review}

Recent advances in generative AI have accelerated the development of semantic communication for future wireless networks. Early studies focused on semantic encoding and deep joint source-channel coding (JSCC), where neural networks were trained to optimize task performance rather than signal reconstruction fidelity \cite{bourtsoulatze2019deep}. More recently, semantic communication has been extended to vehicular and multimodal environments, where camera images, LiDAR point clouds, and textual context are transformed into compact semantic representations for cooperative perception and autonomous driving applications \cite{lu2025generative,xu2023semantic}. These approaches demonstrated significant reductions in communication overhead while preserving task-relevant information. However, most existing frameworks rely on encoder and decoder architectures that transmit semantic representations independently at each time instant, without using higher-level reasoning or coordination among distributed entities.

The emergence of foundation models, particularly LLMs and VLMs, has further expanded the capabilities of semantic communication systems. Recent studies have explored LLM-assisted semantic transmission for context-aware communication management, and multimodal semantic representation learning \cite{salehi2025llm,jiang2025visual,du2025task}. VLMs are particularly attractive for vehicular and sensing applications because they can transform complex visual observations into compact natural-language descriptions while retaining task-relevant semantics \cite{dong2026advances}. Despite these advances, existing work primarily use foundation models for semantic extraction and compression, while providing limited support for cooperative reasoning across multiple agents or shared situational awareness. Recent studies have shown that causal learning can improve robustness, interpretability, and generalization by identifying underlying cause and effect relationships rather than relying solely on statistical correlations \cite{thomas2024causal,thomas2023causal}. In parallel, graph-based world models and neuro-symbolic frameworks have been explored for representing scene dynamics and supporting predictive decision-making in autonomous systems \cite{min2024driveworld}. However, these techniques are rarely integrated into semantic communication architectures, and most existing systems continue to transmit semantic descriptions independently rather than communicating compact updates of shared causal scene representations.

During recent years, agentic AI systems have been used in multi-agent coordination \cite{wei2026agentic} for network management, resource orchestration, and autonomous wireless optimization \cite{dev2025advanced}. However, communication and reasoning are typically treated as separate processes. Existing semantic communication frameworks rarely incorporate autonomous agents that can determine what information should be transmitted, coordinate semantic reporting among distributed entities, adapt transmission fidelity according to scene novelty, or jointly optimize communication and situational awareness objectives \cite{zhang2022deep,yang2022deep}. 
The agentic paradigm introduces autonomous decision-making into the communication process itself \cite{seon2026agentic}. Specifically, vehicle-side agents autonomously evaluate situational novelty to suppress redundant reports and adapt transmission fidelity locally, while an edge RSU agent actively correlates distributed reports to steer shared channel capacity toward emerging hazards \cite{yan2025semantic}. This limitation is particularly important in large-scale vehicular and intelligent sensing environments, where communication resources are limited and many observations are highly predictable or redundant \cite{liu2025collaborative}. Therefore, there remains a significant gap in the literature for a unified framework that combines semantic communication, foundation models, causal reasoning, diffusion-based transmission, and agentic decision-making to support scalable and bandwidth-efficient cooperative perception in next-generation intelligent transportation and sensing systems.

\section{System Model}

In this paper, we consider ARGUS, a hierarchical semantic communication paradigm in which distributed vehicle agents interact with an edge-enabled roadside orchestrator to support safety-critical perception. Specifically, we consider a fleet of $N$ connected autonomous vehicles operating in an urban cellular V2X (C-V2X) environment, where each vehicle $v_i$ is equipped with an onboard camera and an onboard unit (OBU) with integrated computing capabilities for real-time perception and semantic encoding. The vehicles communicate with an RSU equipped with edge computing capabilities, which coordinates network-wide orchestration and semantic fusion. All vehicles share a time-varying uplink channel with a total capacity of $B_{\text{total}}(t)$ 
Vehicles communicate with a single RSU equipped with edge computing capabilities, which coordinates orchestration and semantic fusion across the network; 
However, due to limited bandwidth and dynamic channel conditions caused by mobility, fading, and interference, transmitting raw sensor data from all vehicles is infeasible. Consequently, communication is restricted to task-relevant semantic information while satisfying strict latency and reliability requirements. Accordingly, the system objective is to maximize the utility of the semantic information exchanged under communication, reliability, and bandwidth constraints, while ensuring that safety-critical events are reliably detected and prioritized.

For semantic representation, at each time instant $t$, vehicle $v_i$ acquires a local observation $\mathbf{O}_i(t)$ through its onboard perception system. In the traditional communication paradigm, a compressed representation $\hat{\mathbf{O}}_i(t)$ of this observation is transmitted, requiring $R_i^{\text{trad}}$ bits. In contrast, the proposed semantic communication framework encodes the observation into a compact semantic message as
\begin{equation}
    \mathbf{s}_i(t) = f_{\text{enc}}\left(\mathbf{O}_i(t),\, \mathcal{K}(t)\right),
\end{equation}
where $f_{\text{enc}}(\cdot)$ denotes the semantic encoder and $\mathcal{K}(t)$ represents the shared knowledge base available at time $t$. The resulting semantic representation requires $R_i^{\text{sem}}$ bits for transmission. Upon reception at the RSU, the semantic message is decoded to reconstruct a task-relevant representation of the scene as
\begin{equation}
    \tilde{\mathbf{O}}_i(t) = f_{\text{dec}}\left(\mathbf{s}_i(t),\, \mathcal{K}(t)\right),
\end{equation}
where $f_{\text{dec}}(\cdot)$ denotes the semantic decoder. This enables task-oriented reconstruction of the observed scene while substantially reducing communication overhead compared with conventional communication approaches.

\subsection{Novelty Score}

In this system, the vehicle agent computes a novelty score $\eta_i(t) \in [0,1]$ that quantifies the deviation of the current observation from a per-vehicle running estimate of the expected scene. It is defined as
\begin{equation}
    \eta_i(t) = 1 - \frac{\left\langle \phi\left(\mathbf{O}_i(t)\right),\, \bar{\phi}_i(t) \right\rangle}{\left\|\phi\left(\mathbf{O}_i(t)\right)\right\| \cdot \left\|\bar{\phi}_i(t)\right\|},
\end{equation}
where $\phi(\cdot)$ is the pooled vision-tower embedding generated by the VLM used for captioning, and $\bar{\phi}_i(t)$ is an exponential moving average of vehicle $v_i$'s past embeddings, $\bar{\phi}_i(t) = \delta\,\bar{\phi}_i(t-1) + (1-\delta)\,\phi(\mathbf{O}_i(t))$ with decay $\delta=0.95$, initialized to the first observed embedding. This lightweight running average plays the role of the expected scene rather than a query against a structured, multi-layer knowledge store. 

In ARGUS, the novelty score approaches zero when the observed scene aligns with the vehicle's own recent history and increases toward one as the deviation grows. Accordingly, each vehicle agent selects a transmission rate $r_i(t)$ as a function of the novelty score and a channel-quality estimate $h_i(t) \in [0,1]$, derived from the link SNR and reflecting instantaneous channel conditions: 
\begin{equation}
    r_i(t) = g\left(\eta_i(t),\, h_i(t)\right),
\end{equation}
where the mapping $g(\cdot)$ selects $r_i(t)$ from a discrete set of transmission modes $r_i(t) \in \left\{r_{\text{heartbeat}},\, r_{\text{semantic}},\, r_{\text{thumbnail}},\, r_{\text{patch}},\, r_{\text{full}}\right\}$, with corresponding bit rates satisfying $r_{\text{heartbeat}} \ll r_{\text{semantic}} < r_{\text{thumbnail}} < r_{\text{patch}} \ll r_{\text{full}}$. Accordingly, during routine driving, most vehicles transmit only heartbeat or compact semantic messages, resulting in very high compression ratios. During safety-critical situations, only a subset of vehicles transmit richer semantic representations or detailed observations, reducing the compression ratio while remaining significantly more bandwidth-efficient than continuous raw streaming. 

\subsection{Bandwidth Allocation}
\label{sec:bandwidth-alloc}

We consider a single-cell vehicular network in which one RSU serves all associated vehicles. Accordingly, the uplink bandwidth managed by the RSU is shared among the vehicles within this cell
using a weighted proportional heuristic designed to satisfy three requirements:
\setcounter{equation}{4} 
\begin{subequations}\label{eqn:5}
\begin{align}
    \sum_{i=1}^{N} b_i(t) &\leq B_{\text{total}}(t), \label{eqn:5a} \\
    b_i(t) &\geq b_{\min}, \quad \forall\, i, \label{eqn:5b} \\
    b_i(t) &\geq r_i(t), \quad \text{if } \eta_i(t) > \eta_{\text{critical}}, \label{eqn:5c}
\end{align}
\end{subequations}
where $b_i(t)$ denotes the bandwidth allocated to vehicle $v_i$, $B_{\text{total}}(t)$ is the total available uplink capacity, $b_{\min}$ is the minimum per-vehicle allocation, $r_i(t)$ is the requested rate associated with the vehicle's selected transmission mode, and $\eta_{\text{critical}}$ is a fixed critical novelty threshold. Specifically, \eqref{eqn:5a} ensures that the total allocation does not exceed the available capacity, \eqref{eqn:5b} guarantees that every vehicle receives at least a minimum heartbeat-level allocation, and \eqref{eqn:5c} ensures that any vehicle whose novelty exceeds the critical threshold receives at least its full requested rate.

To satisfy these requirements, the allocator first assigns the full requested rate to vehicles with $\eta_i(t) > \eta_{\text{critical}}$, thereby satisfying \eqref{eqn:5c}. The remaining bandwidth is then distributed among the other vehicles according to the priority weight
\begin{equation}
    w_i(t) = \alpha \cdot \eta_i(t) + \beta \cdot o_i(t) + \gamma \cdot m_i(t), \quad \alpha+\beta+\gamma=1,
\end{equation}
where $o_i(t) \in \{0,1\}$ indicates whether the vehicle is being viewed by an operator, and $m_i(t) \in [0,1]$ denotes its mission-priority weight. The weighting coefficients are set to $\alpha=0.6$, $\beta=0.25$, and $\gamma=0.15$. Each vehicle's proportional allocation is capped at its requested rate. If the resulting total allocation exceeds $B_{\text{total}}(t)$, the allocations are rescaled while preserving the minimum allocation constraint in \eqref{eqn:5a}. Each scheduling interval is initialized by assigning $b_{\min}$ to every vehicle, thereby ensuring the minimum heartbeat-level allocation, ~\eqref{eqn:5b}.

\subsection{Transmission Rate Selection}

Each vehicle agent selects its transmission rate $r_i(t)$ from a discrete set of transmission modes using a deterministic novelty- and channel-aware rule rather than a learned or reinforcement-trained policy. Routine, low-novelty scenes are mapped to lightweight modes, such as heartbeat and semantic caption, whereas increasingly novel scenes are mapped to richer modes, such as thumbnail, patch, and full-stream transmission. When the estimated channel quality is poor, the selected mode is downgraded to reduce the communication load. This deterministic policy provides a practical instantiation of the rate and bandwidth trade-off that could, in principle, be optimized by a learned policy.

Rather than transmitting raw sensory observations, each vehicle sends a compact semantic report consisting of (i) a natural-language description generated by the onboard VLM, (ii) a dense semantic embedding capturing high-level scene features, and (iii) a novelty score indicating the deviation of the current observation from the expected semantic state. These representations enable the RSU to perform cooperative semantic perception while substantially reducing communication overhead. The resulting communication reduction relative to traditional compressed sensory transmission is quantified by the overall compression ratio:
\begin{equation}
    \rho(t) =
    \frac{\sum_{i=1}^{N} R_i^{\text{trad}}}
         {\sum_{i=1}^{N} R_i(t)}
    =
    \frac{N \cdot R^{\text{trad}}}
         {\sum_{i=1}^{N} r_i(t)\, \Delta t},
\end{equation}
where $R_i^{\text{trad}}$ denotes the communication cost of transmitting the compressed raw observation of vehicle $v_i$ under the traditional paradigm, and $R_i(t)$ denotes its semantic communication cost at time $t$, approximated by the selected transmission rate $r_i(t)$ over the schedule interval $\Delta t$.

\section{ARGUS Architecture: VLM-LLM based agentic semantic communication}

\begin{figure*}
    \centering
    \includegraphics[width=0.93\linewidth]{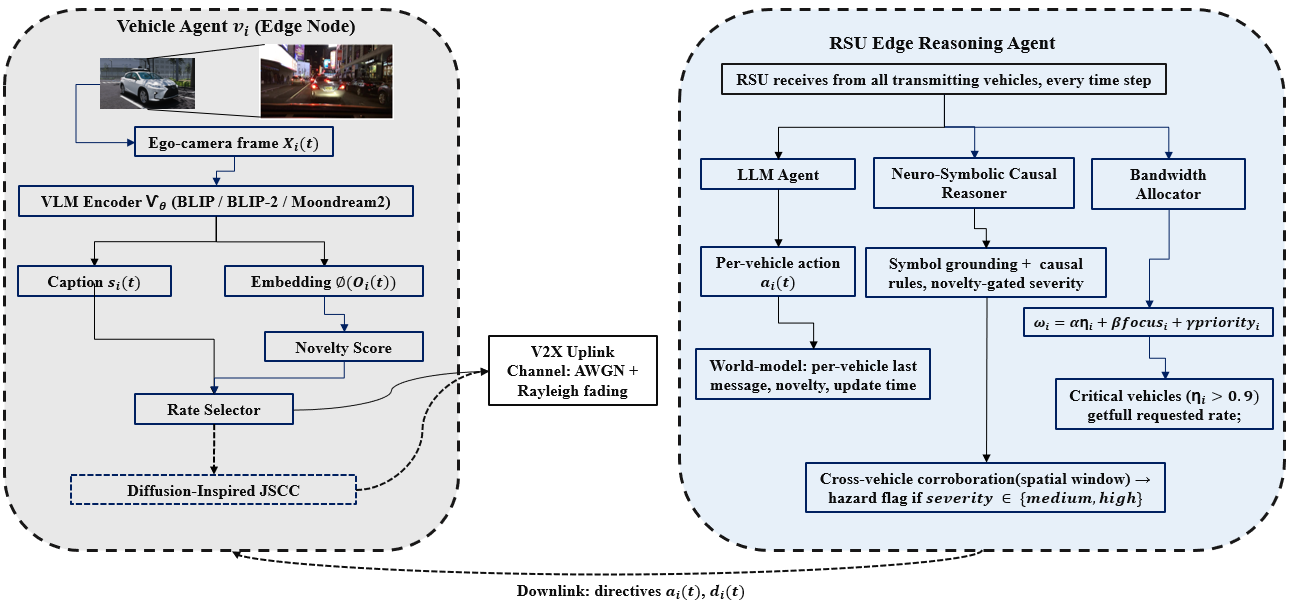}
    \caption{Overall ARGUS system architecture}
    \label{fig:architecture}
            \vspace{-0.5cm}
\end{figure*}

ARGUS comprises of two main components, vehicle-side intelligent agents and an edge reasoning agent deployed at the RSU. Vehicle agents are responsible for local perception, semantic encoding, and rate selection, while the RSU performs semantic reasoning, causal hazard inference, and bandwidth allocation. Rather than a synchronized, richly structured knowledge base, coordination is achieved through the compact semantic messages and a small amount of state kept independently by each side, described below. Fig.~\ref{fig:architecture} summarizes this end-to-end (E2E) architecture and the flow of data between its components.

\subsection{Vehicle Agent and RSU Agent}
\label{sec:vehicle-agent}

Each connected autonomous vehicle is equipped with the OBU 
unit that operates as an intelligent communication agent. The agent continuously processes its camera frame and determines the most informative representation to transmit over the V2X network. It consists of three functional modules: a VLM-based perception module, a novelty scorer, and a rate selector. The VLM-based perception module processes the onboard camera frame using a pretrained VLM to generate a natural-language scene caption along with a pooled vision embedding that encodes the current driving context. The novelty scorer estimates the semantic novelty of the current observation by comparing its embedding against a per-vehicle exponential moving average of its own recent embeddings, a lightweight, local substitute for a predicted expected scene, with no shared or externally supplied prediction. The resulting novelty score reflects how unexpected the observed scene is relative to that vehicle's own recent history. The rate selector then determines the transmission mode from the computed novelty score and the estimated channel quality; this mapping is a fixed threshold rule.
The selected semantic message is then transmitted over the V2X link.

The RSU hosts the edge reasoning agent that receives semantic reports from all vehicles and performs cooperative traffic-event monitoring, semantic reasoning, event assessment, and bandwidth allocation across the participating vehicles. It comprises an LLM-based semantic reasoner, a neuro-symbolic causal reasoner, and a bandwidth allocation agent as three separate modules. At each time step, the LLM-based semantic reasoner processes the incoming vehicle messages in a single prompt. It outputs an action for each vehicle (normal, monitor, request-detail, or elevate) alongside any cross-vehicle event correlations. Additionally, it tracks each vehicle's latest message, novelty score, and timestamp, providing a state representation. The neuro-symbolic reasoner converts incoming semantic descriptions into structured traffic objects and applies causal rules to predict road hazards, adjusting predicted severity based on scene novelty. When neighboring vehicles observe the same scene, their reports are combined to track developing traffic events. These hazard and novelty ratings are then passed to a separate deterministic module that allocates uplink bandwidth based on mission priorities, rather than relying on the LLM to make network decisions. By managing both safety reasoning and bandwidth through direct messages and local logs, the system eliminates the overhead of maintaining synchronized global maps or shared databases across the network. Specifically, ARGUS targets cooperative traffic monitoring scenarios, such as tracking developing congestion, lane blockages, or temporary work zones, where supervisory decisions operate at the scale of seconds and the inference latency of edge LLM reasoning is readily accommodated.

\subsection{Sender Side: VLM-Based Semantic Encoding}
\label{sec:vlm-encoder}

\subsubsection{Model Architecture}

In ARGUS, each connected vehicle $v_i$ hosts an OBU that runs a pretrained VLM, denoted as $\mathcal{V}_\theta$, directly for inference without fine-tuning or distillation. To realize $\mathcal{V}_\theta$, the architecture supports three interchangeable pretrained backends: BLIP, BLIP-2, and Moondream2, each queried with the fixed driving prompt. To generate context-aware representations, this model takes vehicle $v_i$'s camera frame $\mathbf{X}_i(t)$ along with a fixed prompt that directs its attention toward road agents and safety hazards rather than background scenery. From this single conditioned pass, the VLM extracts two complementary outputs: a text description $\mathbf{s}_i(t) = \mathcal{V}_\theta(\mathbf{X}_i(t))$ used for symbolic reasoning and a visual feature vector $\phi(\mathbf{O}_i(t))$ used to measure scene novelty.

\subsection{Receiver Side: LLM-Based Agentic Reasoning}
\label{sec:receiver-side}

On the receiver side, the central LLM agent hosted at the RSU runs a locally hosted LLM (Llama 3.2, via Ollama \cite{grattafiori2024llama}) $\mathcal{L}_\phi$ that serves as the cooperative reasoning engine 
ARGUS provides a modular perception front-end with three interchangeable VLMs tailored to heterogeneous vehicular compute budgets. This enables resource-constrained vehicles to prioritize low-latency inference while allowing higher-compute platforms to extract richer scene semantics, demonstrating that ARGUS is fully model-agnostic. At each time step, it receives the semantic reports $\{(\mathbf{s}_i(t), \eta_i(t))\}_{i=1}^{N}$ from every vehicle that has transmitted, and issues a combined prompted call, that jointly performs per-vehicle assessment and cross-vehicle correlation:
\begin{equation}
\begin{split}
    \left(\{a_i(t), \mathbf{d}_i(t)\}_{i=1}^{N},\, \mathcal{G}(t)\right) = {} & \mathcal{L}_\phi\Big(\left\{\mathbf{s}_i(t), \eta_i(t)\right\}_{i=1}^{N} \\
    & \quad \;\Big|\; \texttt{prompt}_{\text{reason}}\Big),
\end{split}
\end{equation}
where $a_i(t) \in \{normal, monitor, request_{detail}, elevate\}$ is the assessed action level for vehicle $v_i$, $\mathbf{d}_i(t)$ is an accompanying short natural-language rationale, and $\mathcal{G}(t)$ is a set of inferred cross-vehicle event correlations (vehicle-id groups sharing a common reported event, with an associated severity). This natural-language rationale ensures that supervisory directives remain explainable and auditable to human operators, following interpretable autonomous driving frameworks that pair operational actions with linguistic justifications \cite{tian2024drivevlm}. For instance, in scenarios involving road closures or stationary obstacles, $\mathbf{d}_i(t)$ documents the underlying hazard context (e.g., ``lane obstructed ahead, high collision risk'') rather than producing an opaque classification label. If the LLM's response omits a vehicle, or a vehicle has not transmitted recently, a deterministic default directive is used ($normal$, or $request_{detail}$ if the vehicle has been silent for more than 10 time steps) rather than a further LLM call. The agent additionally maintains a simple per-vehicle record of the most recent message, novelty score, and update time; this record is updated deterministically after each call rather than being regenerated by the LLM itself, and it is not shared with or queried by the vehicles 
This short text explanation makes decisions easy for human operators to check and understand, following explainable self-driving systems that pair actions with plain-language reasons \cite{tian2024drivevlm}.

Each vehicle transmits its semantic caption and novelty score over the uplink channel. To evaluate communication overhead without loss of generality, we adopt a discrete payload model based on representative data budgets: 20~B for a heartbeat beacon, 200~B for a compact semantic caption, 5~KB for a thumbnail keyframe, 30~KB for a region-of-interest patch, and 500~KB for a raw compressed frame \cite{etsi2011intelligent}. These benchmarks reflect practical message sizes across multimodal transmission modes without imposing assumptions on a specific text source-coding scheme.

\subsection{Neuro-Symbolic Causal Hazard Reasoning}
\label{sec:causal}

While the VLM--LLM framework communicates semantic descriptions of observed scenes, ARGUS augments this process with a neuro-symbolic causal reasoner that grounds each vehicle's caption in symbolic driving concepts and applies deterministic causal rules to infer potential hazards. Specifically, each caption $\mathbf{s}_i(t)$ is mapped to a set of symbols $S_i(t) \subseteq \Sigma$ through nine regular-expression keyword matching, listed in the Table. \ref{tab:symbol-lexicon} against a fixed lexicon $\Sigma$ of driving concepts, including \texttt{pedestrian}, \texttt{cyclist}, \texttt{stopped\_vehicle}, \texttt{heavy\_vehicle}, \texttt{intersection}, \texttt{low\_visibility}, \texttt{obstacle}, \texttt{collision}, and \texttt{dense\_traffic}. Each concept is associated with a small set of trigger words or phrases; for example, \texttt{pedestrian} is triggered by terms such as pedestrian, person, walker, or jaywalk. This grounding process is entirely deterministic and relies only on the content of the received caption.

\begin{table}[t]
\centering
\footnotesize
\caption{Symbol lexicon $\Sigma$ used for regex-based caption grounding.}
\label{tab:symbol-lexicon}
\setlength{\tabcolsep}{4pt}
\renewcommand{\arraystretch}{0.9}
\begin{tabular}{@{}l p{5.5cm}@{}}
\toprule
\textbf{Symbol} & \textbf{Trigger words / phrases} \\
\midrule
\texttt{pedestrian}      & pedestrian, person, people, walker, jaywalk \\
\texttt{cyclist}         & cyclist, bicycle, biker, motorcycle, motorcyclist \\
\texttt{stopped\_vehicle} & stopped, stalled, parked, halted, braking, brake \\
\texttt{heavy\_vehicle}   & truck, bus, trailer, semi \\
\texttt{intersection}    & intersection, junction, crossroad, crosswalk, traffic light \\
\texttt{low\_visibility}  & night, fog, foggy, dark, rain, rainy, wet, snow, glare \\
\texttt{obstacle}        & debris, obstacle, object, construction, cone, barrier, pothole \\
\texttt{collision}       & accident, collision, crash, wreck, overturned \\
\texttt{dense\_traffic}   & traffic jam, congestion, queue, heavy traffic, many cars \\
\bottomrule
\end{tabular}
        \vspace{-0.3cm}
\end{table}

The grounded symbols are then evaluated against a predefined knowledge base containing eight deterministic causal rules, as listed in Table. \ref{tab:causal-rules}. Each rule maps a combination of symbolic entities to a hazard type, an intrinsic baseline severity, and a recommended mitigation action. For example, the $\text{lead\_vehicle\_braking}$ rule is triggered when $\text{stopped\_vehicle} \in S_i(t)$ and infers a rear-end collision risk with a $\text{high}$ baseline severity. In contrast, the $\text{vulnerable\_road\_user\_crossing}$ rule additionally requires $\text{intersection} \in S_i(t)$. More generally, a rule is activated for vehicle $i$ when its required antecedent predicates, or a valid disjunctive subset, are satisfied within $S_i(t)$, allowing multiple rules to be triggered concurrently.

\begin{table*}[!t]
\centering
\footnotesize
\caption{Causal rule knowledge base used for symbolic hazard inference.}
\label{tab:causal-rules}
\setlength{\tabcolsep}{5pt}
\renewcommand{\arraystretch}{0.95}
\begin{tabular}{@{}l p{4.5cm} c p{4.5cm}@{}}
\toprule
\textbf{Rule name} & \textbf{Antecedent(s)} & \textbf{Severity} & \textbf{Action} \\
\midrule
\texttt{vulnerable\_road\_user\_crossing} & \texttt{pedestrian} \texttt{intersection}) & high & \texttt{prepare\_to\_yield\_and\_brake} \\
\texttt{vru\_present}                     & \texttt{pedestrian}        & medium & \texttt{reduce\_speed} \\
\texttt{cyclist\_share\_lane}             & \texttt{cyclist}           & medium & \texttt{increase\_lateral\_gap} \\
\texttt{lead\_vehicle\_braking}           & \texttt{stopped\_vehicle}  & high   & \texttt{increase\_following\_distance} \\
\texttt{road\_obstacle}                   & \texttt{obstacle}          & high   & \texttt{plan\_avoidance\_maneuver} \\
\texttt{crash\_ahead}                     & \texttt{collision}         & high   & \texttt{slow\_and\_reroute} \\
\texttt{degraded\_conditions}             & \texttt{low\_visibility}   & medium & \texttt{extend\_stopping\_distance} \\
\texttt{congestion}                       & \texttt{dense\_traffic}    & low    & \texttt{maintain\_gap} \\
\bottomrule
\end{tabular}
        \vspace{-0.3cm}
\end{table*}

The resulting hazard assessments are subsequently adjusted according to the novelty of the current observation. When $\eta_i(t)$ exceeds the escalation threshold, set to $0.6$ in the ARGUS implementation, the inferred severity is increased by one grade along the ordinal scale ${\text{none} < \text{low} < \text{medium} < \text{high}}$, with $\text{high}$ serving as the maximum severity level. 
This novelty adjustment prioritizes hazards deviating significantly from expectation.

To incorporate corroborating evidence from multiple vehicles, predictions of the same hazard type that are independently triggered by different vehicles within a predefined spatial radius are grouped into spatial clusters. When a cluster contains multiple reporting vehicles, the corresponding hazard is classified as propagating, and the severity of each contributing prediction is increased by one grade. Importantly, this cross-vehicle corroboration only modifies the severity of hazards that have already been independently identified from a vehicle's own observation; it does not generate a new hazard alert for an otherwise unobservant vehicle solely from neighboring reports.

Finally, vehicle $v_i$ is classified as hazardous if any of its active rules, after novelty-based adjustment and spatial corroboration, reaches a $\text{medium}$ or $\text{high}$ severity level. 
Thus, per-step hazard classification relies solely on grounded symbols, novelty metrics, and neighboring semantic reports. Vehicles transmit only compact semantic captions, leaving causal inference, hazard evaluation, and corroboration to the RSU. Consequently, ARGUS eliminates the overhead of sending explicit causal graphs or predictive models.

\subsection{Diffusion-Inspired Semantic Transceiver}
\label{sec:diffusion}

While Shannon's source-channel separation theorem guarantees optimality under infinite block lengths, it exhibits sub-optimal performance for the short, variable-length semantic representations characteristic of vehicular networks \cite{polyanskiy2010channel}. To overcome this limitation, we adopt a JSCC framework for semantic embeddings inspired by denoising diffusion probabilistic models \cite{bourtsoulatze2019deep}. By drawing a structural equivalence between physical channel corruption and the forward diffusion process, the downstream semantic recovery can be naturally formulated and solved as a reverse denoising process.

Given an embedded semantic representation $\mathbf{x}_i^{(0)} = \phi(\mathbf{O}_i(t))$, the forward process adds noise over a fixed linear schedule $\{\beta_k\}_{k=1}^{K}$, where $\bar{\alpha}_k = \prod_{j=1}^{k}(1-\beta_j)$ scales the original signal:
\begin{equation}
    \mathbf{x}_i^{(k)} = \sqrt{\bar{\alpha}_k}\, \mathbf{x}_i^{(0)} + \sqrt{1 - \bar{\alpha}_k}\, \boldsymbol{\epsilon}_k, \quad \boldsymbol{\epsilon}_k \sim \mathcal{N}(\mathbf{0}, \mathbf{I}).
\end{equation}
where $K=50$ and the variance schedule $\beta_k$ increases linearly from $10^{-4}$ to $0.02$. The diffusion depth $k^*$ is determined deterministically from the channel conditions via a fixed linear mapping, $k^* = \mathrm{clip}(K - 2\cdot\mathrm{SNR}_{\text{dB}},\, 5,\, K-1)$, which assigns a larger perturbation index to lower SNRs. The corrupted representation $\mathbf{x}_i^{(k^*)}$ is transmitted across the physical channel, and upon reception, an analytical minimum mean square error (MMSE)-style reverse update, rather than a learned denoising network $\boldsymbol{\epsilon}_\xi$, is executed from $k^*$ down to $0$ to reconstruct $\hat{\mathbf{x}}_i^{(0)}$, followed by $\ell_2$-normalization to yield $\hat{\mathbf{s}}_i(t)$. By recovering the semantic embedding through progressive noise removal rather than requiring error-free digital packet delivery, this framework mitigates the classical cliff effect: at low SNRs, the receiver reliably retains a coarse semantic representation instead of suffering catastrophic outage. Consequently, this training-free design inherently guarantees graceful degradation without requiring E2E parameter optimization across the transceiver.

\subsection{Agent Decision-Making Process}

Each vehicle agent operates a continuous perception-action loop at frame rate consisting of observation, assessment, and transmission decision phases. First, during observation, the onboard VLM processes the current ego-camera frame to generate both a natural-language caption and a pooled visual embedding, entirely bypassing standalone object-detection or multi-object-tracking pipelines in favor of E2E VLM scene interpretation. Next, the agent assesses the scene by computing a novelty score that quantifies the divergence between the current embedding and its recent historical trajectory. Finally, based jointly on this novelty metric and the prevailing channel conditions, the agent deterministically selects an appropriate transmission mode:

\begin{itemize}
    \item \textbf{Novelty $<0.1$:} Transmit a heartbeat message indicating that the vehicle is operating normally and the observed scene matches the expected semantic context.
    \item \textbf{Novelty $0.1-0.3$:} Transmit a compact semantic report describing the current scene.
    \item \textbf{Novelty $0.3-0.6$:} Transmit the semantic report together with low-resolution visual evidence (thumbnail).
    \item \textbf{Novelty $0.6-0.8$:} Transmit the semantic report together with a high-information image patch.
    \item \textbf{Novelty $>0.8$:} Request full semantic stream transmission for safety-critical situations.
\end{itemize}

The selected semantic representation is subsequently transmitted over a simulated C-V2X channel subject to Rayleigh fading and additive white Gaussian noise (AWGN), capturing both the non-line-of-sight (NLOS) multipath propagation characteristics of dense vehicular environments and the thermal noise inherent to the receiver. When activated, the diffusion-inspired transceiver encloses this channel transition with its forward perturbation and analytical reverse decoding steps.

Operating at the network edge, the central RSU intelligence agent orchestrates cooperative perception and dynamic resource management across three unified processing stages: edge LLM reasoning, neuro-symbolic causal evaluation, and bandwidth allocation. Upon collecting uplink semantic reports from all connected vehicles, an edge LLM synthesizes the distributed inputs to maintain high-level situational awareness and assign per-vehicle action directives. In parallel, a deterministic neuro-symbolic reasoner grounds natural-language captions into explicit driving concepts and evaluates them against predefined causal safety rules. For example, if a vehicle reports a sudden obstacle ahead, the engine maps this textual description to a critical collision risk rule and assigns it a high-severity hazard level. When a neighboring vehicle independently reports the same obstacle, the system fuses both observations into a single corroborated, propagating hazard with elevated confidence.
Finally, the bandwidth allocation module translates this corroborated hazard severity and each vehicle's novelty score into transmission priorities, dynamically assigning a larger share of the shared C-V2X channel capacity to the vehicles witnessing the hazard while throttling routine reports across the rest of the fleet.

\subsection{RSU Supervisory Directives (Action Vocabulary)}

Building on the semantic reports received from the vehicles, the RSU's LLM-based agent assigns each vehicle an operational directive from a predefined four-action vocabulary. At each time step, the agent assesses the available semantic information and assigns one of the following labels to characterize the situation reported by each vehicle:
\begin{itemize}
    \item \textbf{Normal:} Indicates that the received report does not raise an elevated concern and that no additional supervisory attention is required based on the current assessment.
    \item \textbf{Monitor:} Indicates that the reported situation is potentially ambiguous or may evolve over time, warranting continued attention to subsequent observations.
    \item \textbf{Request Detail:} Incomplete or uncertain reports where additional descriptive detail or visual evidence would be required to resolve scene ambiguity.
    \item \textbf{Elevate:} Indicates that the reported information suggests a potentially safety-critical situation requiring a high-priority assessment.
\end{itemize}

This vocabulary serves as a one-way supervisory assessment rather than a two-way negotiation protocol. The RSU assigns and logs these directives alongside their natural-language rationales to provide an explainable, human-auditable record of edge decisions \cite{tian2024drivevlm, xu2024drivegpt4}; it does not transmit them back to alter the vehicles' onboard rate-selection decisions. It should be noted that, the edge LLM and the neuro-symbolic causal reasoner operate in parallel on the incoming reports rather than in a sequence. For example, if a vehicle reports a stationary car ahead, both modules process the caption independently, the causal reasoner evaluates predefined collision rules to determine hazard severity, while the LLM generates the corresponding \textbf{Elevate} directive and text rationale.

\section{Simulation Results}

In this work, we utilized the BDD100K driving dataset \cite{yu2020bdd100k}, a large-scale autonomous driving dataset consisting of approximately 100,000 RGB road scene real-world driving scene images captured from front-facing vehicle-mounted cameras under diverse environmental conditions, including variations in weather, lighting, and urban traffic complexity. The dataset is streamed online and dynamically allocated to a fleet of $N$ vehicles over $T$ time steps, resulting in a structured grid of $N \times T$ observations per simulation episode. Specifically, each vehicle is assigned a temporally ordered sequence of real driving frames, enabling evaluation of ARGUS under realistic visual sensing conditions.

ARGUS needs a locally deployable, relatively lightweight, instruction-following model with sufficient reasoning capability, and Meta Llama 3.2 provides a practical capability–latency–resource trade-off \cite{grattafiori2024llama}. Moreover, 3B Meta Llama 3.2 model outperforms Gemma 2 2.6B and Phi 3.5-mini on instruction following, summarization, prompt rewriting, and tool-use tasks, while its quantized variants provide substantially reduced model size and memory requirements together with 2--4$\times$ faster inference, making it a suitable choice for the latency- and resource-constrained edge reasoning required by ARGUS \cite{grattafiori2024llama}. 
The model is served through the Ollama runtime, enabling low-latency inference at the network edge without reliance on external cloud application programming interface (APIs). The LLM functions as a cooperative reasoning engine that performs two core tasks in a single combined call per time step: (i) interpretation of per-vehicle semantic scene descriptions into an assessed action level, and (ii) cross-vehicle correlation to identify shared or overlapping environmental hazards observed from different viewpoints;

At the vehicle edge, we employ VLMs, including BLIP \cite{li2022blip}, BLIP-2 \cite{li2023blip2}, and Moondream2 \cite{vikhyat2024moondream2}, selected according to available computational resources and task requirements. BLIP is suited to resource-constrained vehicles requiring low-latency captioning, with the smallest parameter count of the three ($\sim$250M) and no LLM decoder. Moondream2 targets moderate-compute platforms where hazard recall is prioritized; it uses visual question answering with a hazard-specific query rather than free-form captioning, at a moderate computational cost ($\sim$1.8B parameters). BLIP-2 is intended for compute-rich platforms where richer captions are preferred, using the OPT-2.7B LLM decoder at higher latency and memory cost. Each VLM produces two complementary outputs: (i) a natural-language semantic message describing the scene, road participants, and potential hazards, and (ii) a dense semantic embedding extracted from the vision backbone to represent high-level visual features in a continuous vector space.

\subsection{Evaluation Framework}

To ensure a fair comparison between ARGUS and the baseline methods, we adopt a unified communication interface in which all methods receive the same vehicle observations and, at each simulation time step, produce two outputs: the transmitted data volume for each vehicle and the set of vehicles identified as containing hazardous events. This ensures that performance differences arise from the underlying communication and reasoning strategies rather than differences in input data. Each observation includes a VLM-generated semantic novelty score, textual scene description, and ground-truth hazard label. Scene severity is categorized as none, low, medium, or high, with only medium- and high-severity events treated as positive hazard events during evaluation.

Communication overhead is calculated using the predefined payload sizes associated with each transmission mode and baseline implementation: heartbeat (20~B), semantic description (200~B), thumbnail (5~KB), image patch (30~KB), and compressed full video frame (500~KB) \cite{etsi2011intelligent}. In addition, each method specifies normalized computational costs for vehicle-side (edge) processing, RSU-side (central) processing, and communication. These costs account for processing such as VLM inference at the vehicle and semantic or video processing at the RSU, enabling consistent comparison of bandwidth consumption, hazard detection performance, computational complexity, and overall communication efficiency under identical simulation conditions.

\subsubsection{Raw Sensor Streaming}

This baseline represents the conventional V2X communication paradigm, in which every vehicle continuously transmits a compressed full video frame to the RSU without semantic encoding. In our implementation, each vehicle transmits a 500~KB frame at every communication interval, and the RSU declares a hazard when the novelty score received exceeds $\tau_n=0.6$. The approach requires communication complexity of $O(NR_{\mathrm{full}})$ and edge processing complexity of $O(NP)$, where $R_{\mathrm{full}}$ denotes the payload size in full-frame and $P$ is the number of pixels processed per frame. Since each vehicle independently transmits and evaluates its observation, this baseline does not exploit semantic compression, adaptive transmission, or cross-vehicle reasoning.

\subsubsection{Fixed-Rate Semantic Communication}

This baseline uses a VLM at each vehicle to convert captured images into semantic descriptions before transmission, but maintains a fixed communication rate regardless of scene novelty or importance. Each vehicle transmits a 200~B semantic message at every communication interval and a 5~KB thumbnail keyframe every five intervals. Hazard detection is performed independently at the RSU by comparing the received novelty score against $\tau_n=0.6$. The resulting vehicle-side computational complexity is $O(NS)$, where $S$ denotes the number of semantic tokens generated by the VLM. Although this approach substantially reduces bandwidth compared with raw video streaming, it does not adapt the transmission rate to scene dynamics or exploit cross-vehicle collaboration and causal reasoning.

\subsubsection{Novelty-Thresholded SemCom}

This baseline employs event-driven semantic communication, where each vehicle performs local VLM-based perception and transmits additional information only when the observed novelty exceeds a predefined threshold. In our implementation, vehicles transmit a 20~B heartbeat when $\eta_i<0.6$ and a 30~KB region-of-interest image patch when $\eta_i\geq0.6$, with the received patch treated as a potential hazard at the RSU. The vehicle-side computational complexity is $O(NS)$, while the communication complexity is $O(NpR_{\mathrm{patch}})$, where $p$ is the fraction of active vehicles and $R_{\mathrm{patch}}$ is the patch payload size. This strategy reduces communication during routine scenes but remains reactive and does not incorporate cooperative or causal reasoning.

\subsubsection{AgentComm}

This baseline follows the AgentComm framework, in which an LLM extracts task-relevant information and an importance-aware transmission mechanism adapts the transmitted payload while suppressing information that can be inferred from a shared knowledge base. We adapt AgentComm to the per-vehicle uplink, without cross-vehicle correlation or causal reasoning. In our implementation, each vehicle transmits a 20~B heartbeat when $\eta_i<0.2$, a 120~B LLM-extracted message for moderate novelty, and a 5~KB importance-protected payload when $\eta_i\geq0.6$. The communication complexity is $O(NR_{\mathrm{extract}})$, while the computational complexity is $O(NS_{\mathrm{llm}})$, where $S_{\mathrm{llm}}$ denotes the number of tokens processed by the LLM extractor. Thus, AgentComm provides adaptive and knowledge-aware transmission but remains limited to independent per-vehicle importance assessment.

\subsubsection{LMM Vehicle SemCom}

This baseline follows an LMM-based task-oriented semantic communication paradigm in which a VLM adaptively divides each onboard image into patches and assigns an importance score based on visual and semantic salience. In our implementation, the importance score is computed as a weighted combination of the novelty score (70\%) and a lightweight semantic-concept matching score (30\%), and each vehicle transmits up to 10 importance-selected image patches, each with a payload of 300~B, with the number of transmitted patches increasing with the importance score. A potential hazard is declared when the importance score exceeds 0.55. The resulting communication complexity is $O(N k(\mathrm{imp})R_{\mathrm{patch}})$, where $k(\mathrm{imp})$ denotes the importance-dependent number of transmitted patches and $R_{\mathrm{patch}}$ is the patch payload size, while the computational complexity is $O(NS_{\mathrm{lmm}})$, where $S_{\mathrm{lmm}}$ represents the LMM processing cost. Unlike ARGUS, this baseline does not incorporate explicit symbolic hazard reasoning, cross-vehicle corroboration, or diffusion-based joint source-channel coding.

\subsubsection{Unified Multi-Task Semantic Communication (U-DeepSC)}

This baseline follows the unified multi-task semantic communication paradigm (U-DeepSC) \cite{zhang2024unified}, in which a Transformer-based semantic encoder maps each vehicle observation to a fixed-size task-oriented feature representation, which is transmitted to the RSU for downstream hazard classification without reconstructing the original image or semantic caption. In our implementation, each vehicle transmits a fixed 1400~B semantic feature at every communication interval, and the RSU employs a Transformer-based decoder and multi-task classification head that combines semantic-concept information with the novelty score to identify potential hazards. The communication complexity is $O(NR_{\mathrm{task}})$, where $R_{\mathrm{task}}$ denotes the fixed feature payload size, while the computational complexity is $O(NS)$, where $S$ represents the semantic feature dimension processed by the decoder and classification head. Although U-DeepSC provides task-oriented semantic transmission without periodic keyframes, it performs hazard recognition independently for each vehicle and does not incorporate cross-vehicle corroboration, causal reasoning, or diffusion-based joint source-channel coding.

\subsection{Communication Efficiency}

\begin{figure*}[t]
\centering
\begin{subfigure}{0.45\textwidth}
\centering
\includegraphics[width=\textwidth]{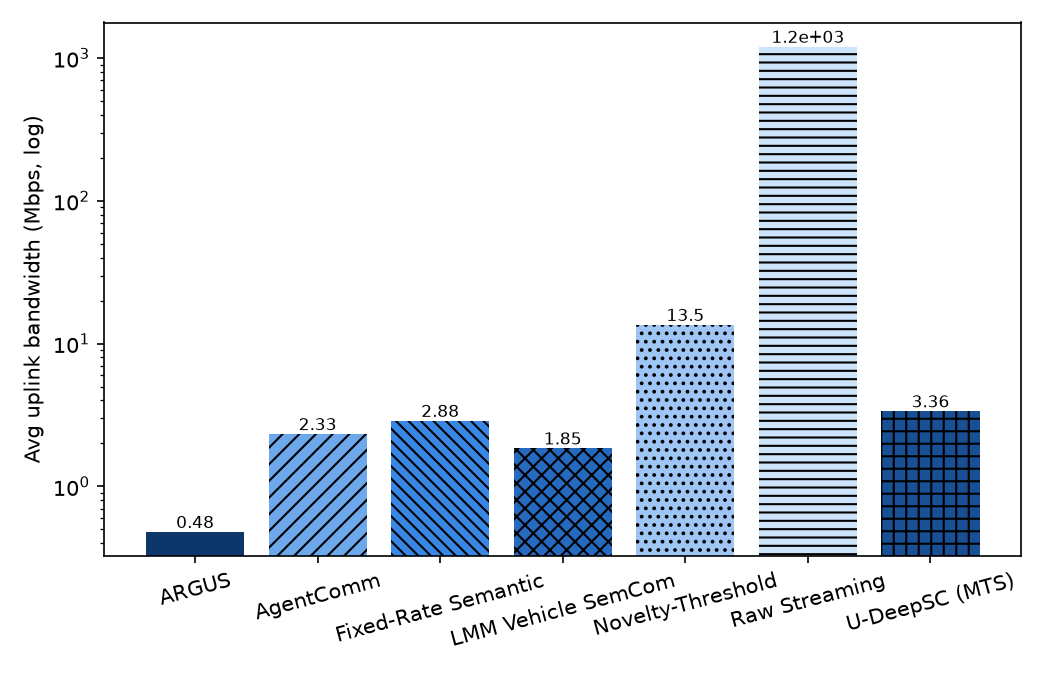}
\caption{Average uplink bandwidth.}
\label{fig:BW}
\end{subfigure}
\hfill
\begin{subfigure}{0.45\textwidth}
\centering
\includegraphics[width=\textwidth]{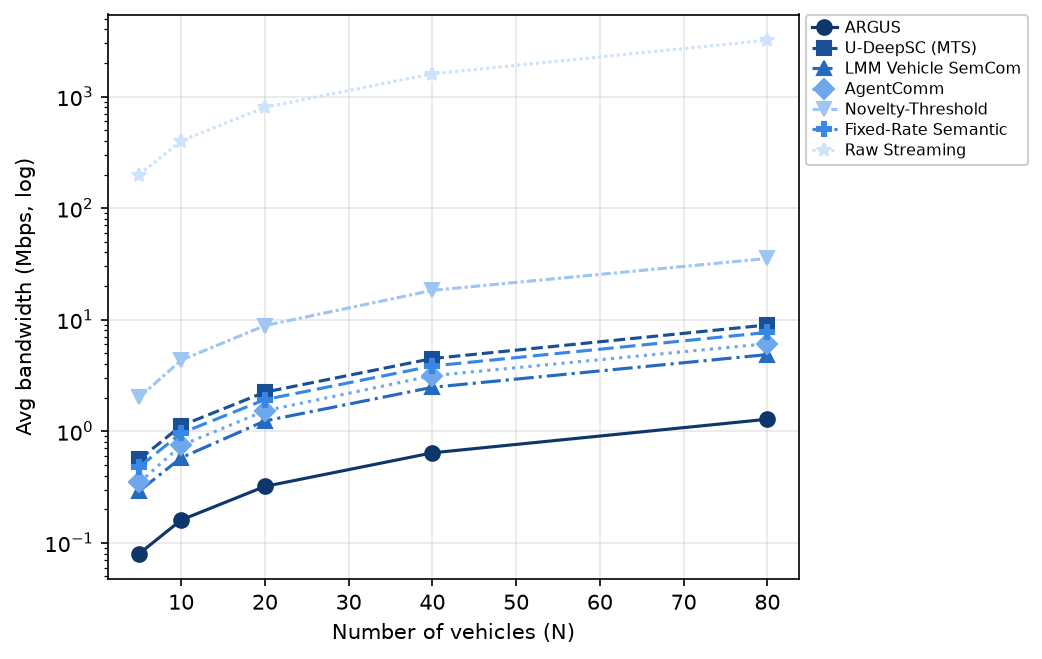}
\caption{Average uplink bandwidth versus fleet size for each method.}
\label{fig:scaling}
\end{subfigure}
\\[1ex]
\begin{subfigure}{0.45\textwidth}
\centering
\includegraphics[width=\textwidth]{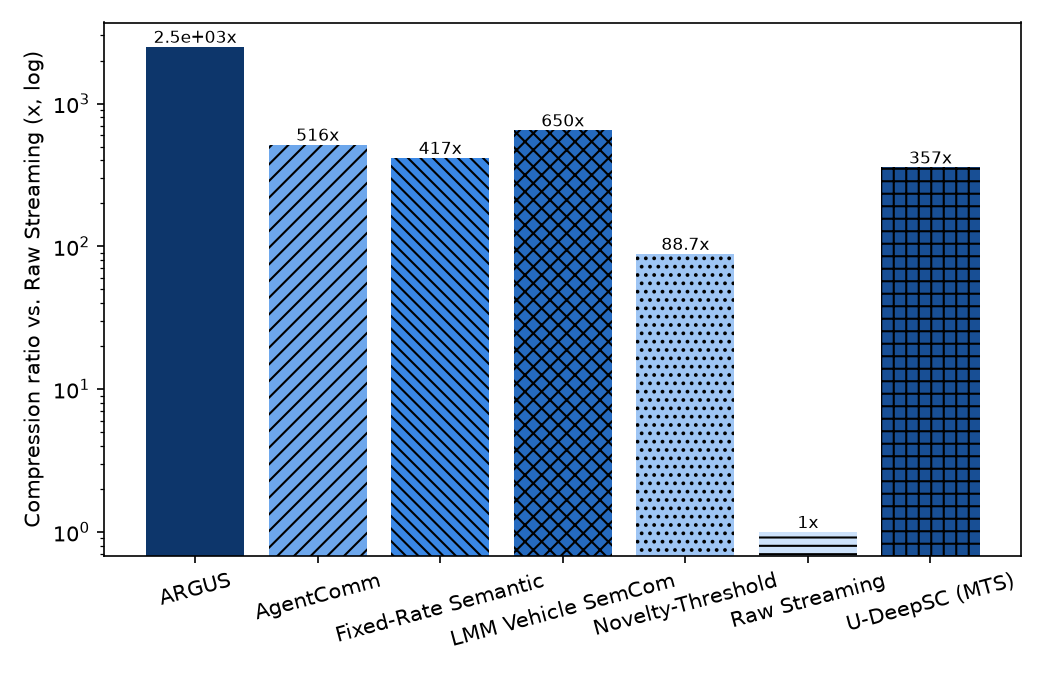}
\caption{Bandwidth reduction factor to Raw Sensor Streaming.}
\label{fig:compression}
\end{subfigure}
\hfill
\begin{subfigure}{0.45\textwidth}
\centering
\includegraphics[width=\textwidth]{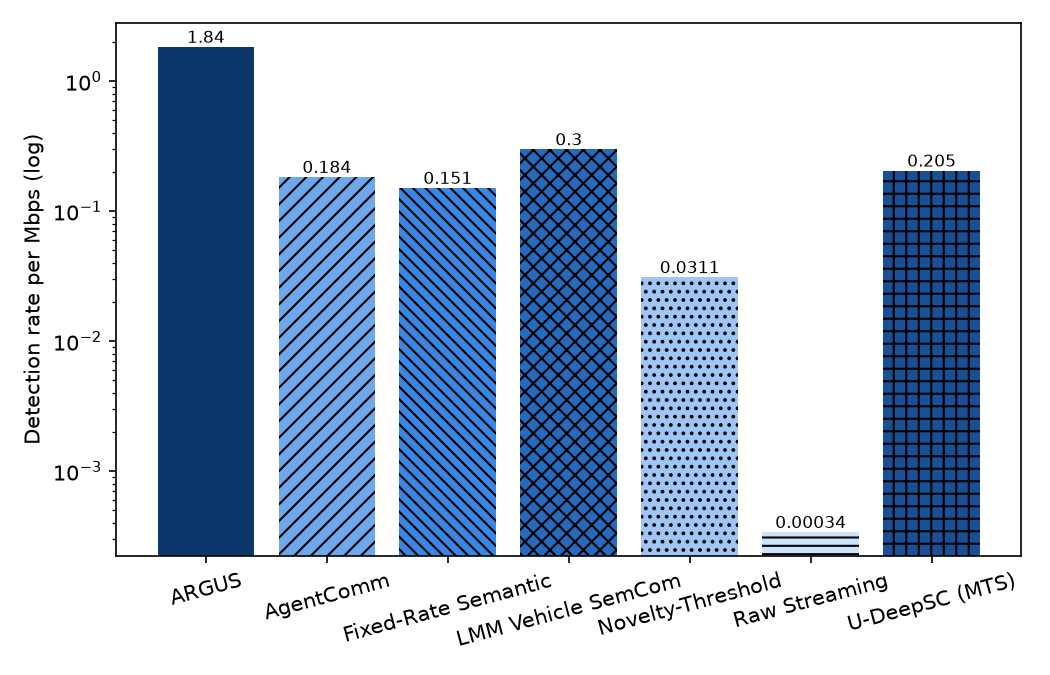}
\caption{Hazard detection rate per Mbps of average bandwidth.}
\label{fig:effmbps}
\end{subfigure}
\caption{Communication efficiency for bandwidth consumption, fleet-size scaling, bandwidth reduction, and detection efficiency.}
\label{fig:BW_scaling_compression_effmbps}
        \vspace{-0.4cm}
\end{figure*}

Communication efficiency characterizes the amount of information required to support semantic perception and hazard detection under the available wireless resources. It is reflected through the average communication load, bandwidth reduction, scaling with the number of vehicles, and the detection performance obtained per unit of bandwidth. Fig.~\ref{fig:BW} compares the average uplink bandwidth of the evaluated methods. ARGUS requires only approximately 0.48~Mbps, substantially reducing communication overhead through compact semantic representations and novelty-aware transmission. LMM Vehicle SemCom and AgentComm require 1.85 and 2.33~Mbps, respectively, while U-DeepSC requires 3.36~Mbps due to its fixed-size semantic representation. Novelty-Thresholded SemCom reaches approximately 13.5~Mbps because novelty events trigger relatively large image-patch transmissions. Raw Sensor Streaming requires approximately 1.2~Gbps due to continuous transmission of full video frames. These results demonstrate that combining semantic compression with adaptive transmission is critical for limiting uplink traffic.

Fig.~\ref{fig:scaling} investigates how the communication requirement changes as the fleet size increases from 5 to 80 vehicles. All approaches exhibit an approximately linear increase, as expected for systems in which vehicles independently contribute to the aggregate uplink load. The difference lies primarily in the per-vehicle communication requirement. ARGUS increases from approximately 0.08~Mbps at five vehicles to 1.28~Mbps at 80 vehicles, whereas the corresponding increase is considerably larger for the other approaches. In particular, Raw Sensor Streaming grows from approximately 200~Mbps to 3.2~Gbps, while Novelty-Thresholded SemCom reaches approximately 35.4~Mbps at 80 vehicles. Thus, although the aggregate load remains dependent on fleet size, the substantially smaller per-vehicle semantic payload allows ARGUS to maintain a lower uplink requirement as the network becomes denser.

Fig.~\ref{fig:compression} expresses the bandwidth savings relative to Raw Sensor Streaming. ARGUS achieves approximately a $2500\times$ reduction in communication, whereas the semantic baselines provide smaller reductions. The difference is not solely due to semantic encoding: ARGUS additionally limits when richer information is transmitted according to scene novelty and therefore avoids repeatedly sending redundant visual content.

Fig.~\ref{fig:effmbps} relates detection performance directly to communication expenditure. ARGUS achieves approximately 1.84 detected hazards per Mbps, considerably higher than the semantic baselines. This gap arises because ARGUS simultaneously reduces the transmitted payload and maintains a high hazard-detection rate. By comparison, Novelty-Thresholded SemCom achieves only approximately 0.031 detections/Mbps, illustrating the cost of transmitting relatively large patches for novelty-triggered events. Raw Sensor Streaming reaches only approximately 0.0003 detections/Mbps in the implemented evaluation, demonstrating that high raw-data throughput does not necessarily translate into proportional task-level communication efficiency.

\subsection{Semantic and Hazard Detection Performance}

\begin{figure*}[t]
\centering
\begin{subfigure}{0.45\textwidth}
\centering
\includegraphics[width=\textwidth]{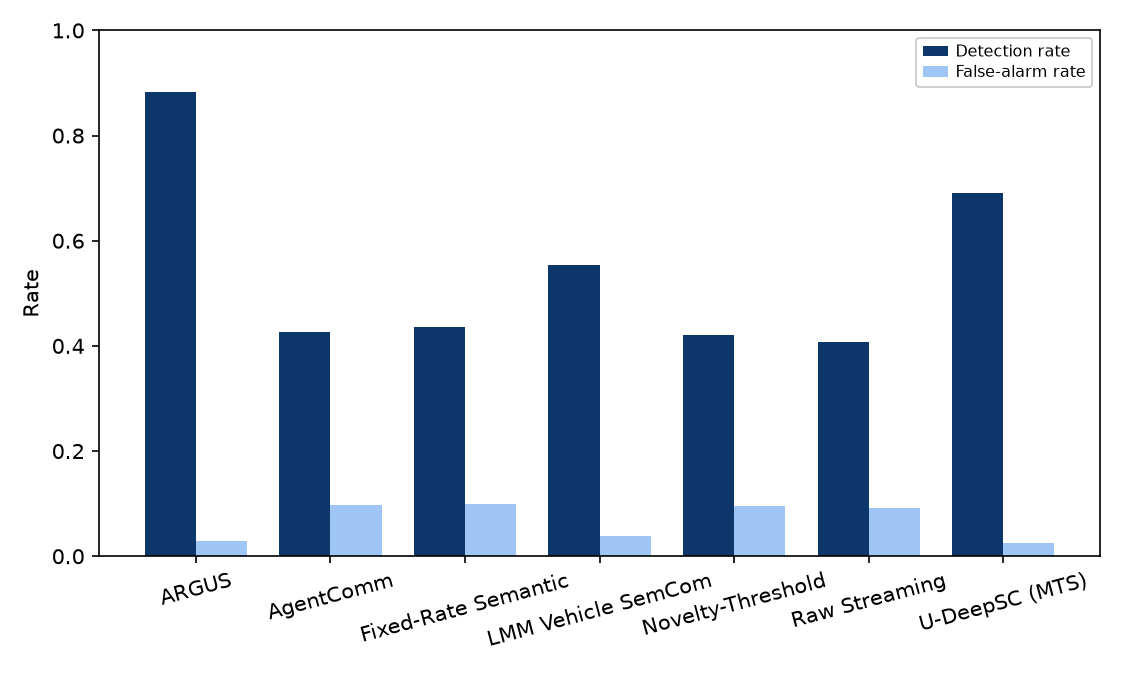}
\caption{Hazard detection rate and FA rate for each method.}
\label{fig:FA}
\end{subfigure}
\hfill
\begin{subfigure}{0.5\textwidth}
\centering
\includegraphics[width=\textwidth]{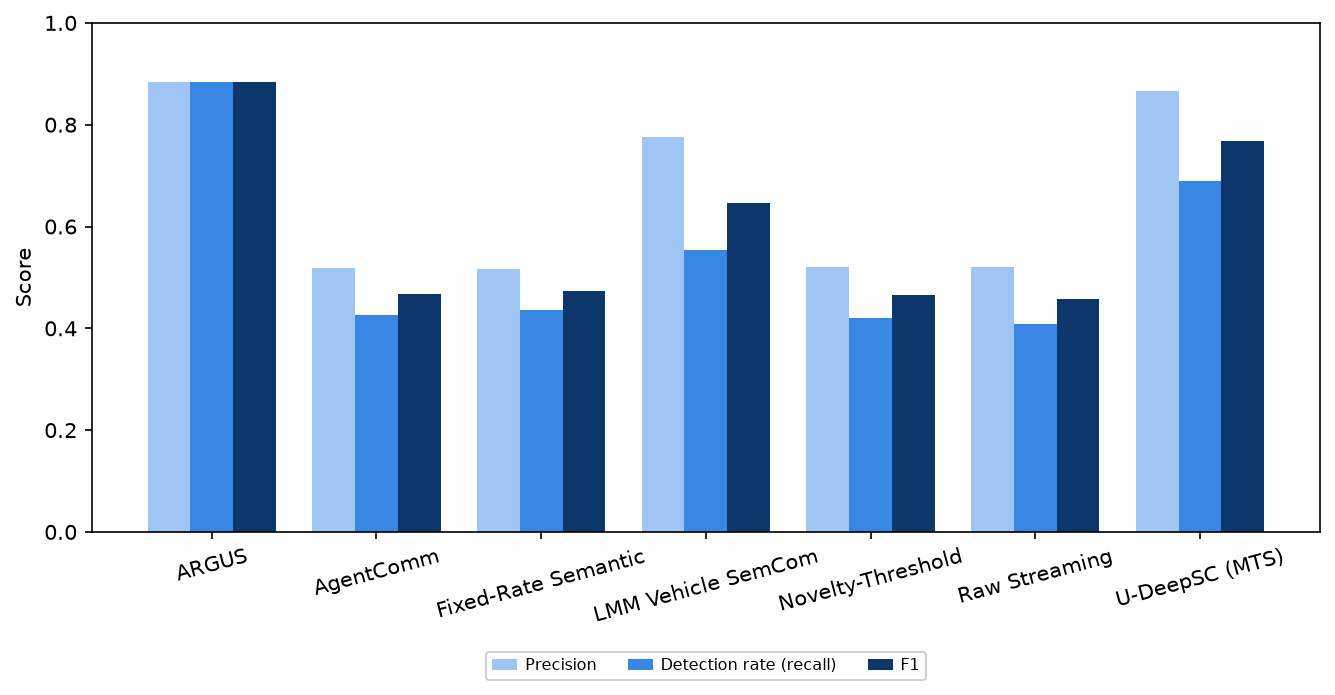}
\caption{Precision, recall (detection rate), and F1 per method.}
\label{fig:prf1}
\end{subfigure}
\\[1ex]
\begin{subfigure}{0.47\textwidth}
\centering
\includegraphics[width=\textwidth]{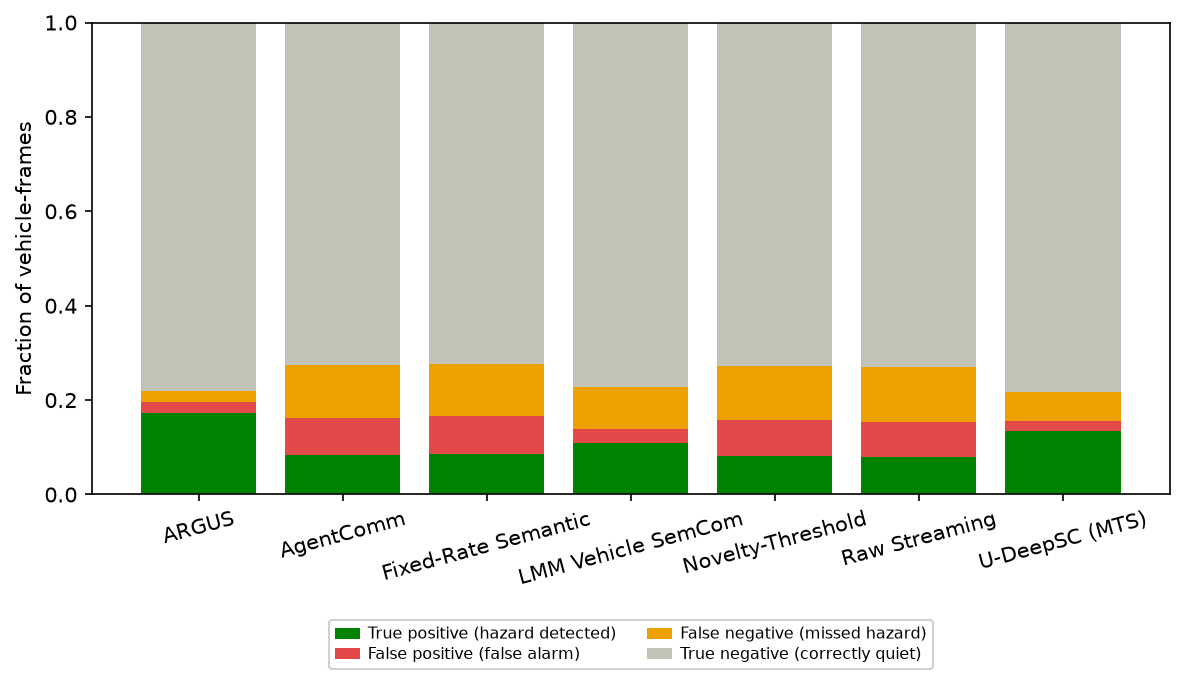}
\caption{Normalized true/false positive/negative composition.}
\label{fig:confusion}
\end{subfigure}
\hfill
\begin{subfigure}{0.45\textwidth}
\centering
\includegraphics[width=\textwidth]{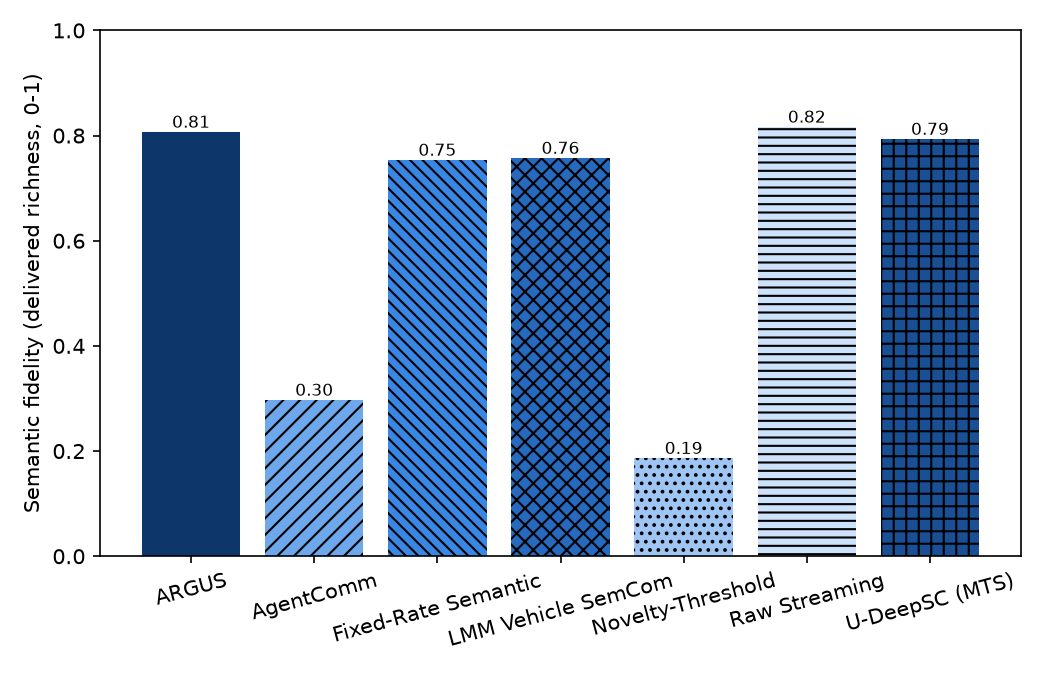}
\caption{Semantic fidelity proxy based on delivered content richness.}
\label{fig:fidelity}
\end{subfigure}
\caption{Semantic and hazard detection performance.}
\label{fig:FA_prf1_confusion_fidelity}
        \vspace{-0.4cm}
\end{figure*}

Task-level performance is characterized by the ability to identify safety-critical events while limiting incorrect hazard decisions. In addition to the overall detection rate and false alarm (FA) rate, precision, recall, F1-score, confusion components, and semantic fidelity provide complementary measures of detection quality and preservation of task-relevant information. Fig.~\ref{fig:FA} compares hazard detection and FA rates. ARGUS achieves a detection rate of 0.883 while maintaining a FA rate of 0.028. This improvement is enabled by combining vehicle-level semantic perception with RSU-side symbolic reasoning and cross-vehicle corroboration. U-DeepSC reaches 0.690 detection with a comparable FA rate of 0.026, whereas the remaining baselines achieve lower detection rates. Notably, Raw Sensor Streaming reaches only 0.408 despite its substantially higher communication cost, indicating that transmitting more raw data does not necessarily improve task-level detection under the adopted decision mechanism.

Fig.~\ref{fig:prf1} provides a more detailed view of hazard-detection quality through precision, recall, and F1-score. ARGUS obtains 0.883 for all three measures, indicating that its high detection rate is not achieved by substantially increasing FA. U-DeepSC obtains a precision of 0.866 and recall of 0.690, resulting in an F1-score of 0.768. LMM Vehicle SemCom provides lower recall despite maintaining a relatively high precision. The lower recall of the remaining baselines indicates that their transmission policies can discard information that is useful to identify hazardous events, particularly when the scene does not trigger their transmission conditions.

Fig.~\ref{fig:confusion} provides the normalized confusion components underlying the detection results. ARGUS produces the largest true-positive fraction (0.173) and the smallest false-negative fraction (0.023), while U-DeepSC obtains corresponding values of 0.135 and 0.061. The larger false-negative observed for the remaining methods are consistent with their lower recall. These results indicate that overall detection differences stem primarily from retaining and correctly identifying hazardous observations rather than from an increase in false positives.

Fig.~\ref{fig:fidelity} compares the semantic fidelity proxy across the evaluated methods. Raw Sensor Streaming obtains the highest value (0.817), consistent with its transmission of substantially richer visual information. ARGUS follows closely at 0.807, indicating that its aggressive reduction in communication does not result in a proportional loss of task-relevant semantic information. U-DeepSC also maintains relatively high fidelity, whereas the lower values observed for AgentComm and Novelty-Thresholded SemCom are consistent with their frequent use of very small heartbeat or event-driven payloads. The marginal fidelity gap between Raw Sensor Streaming and ARGUS is particularly notable alongside the massive reduction in communication overhead.

\subsection{Robustness and Fleet Scalability}

\begin{figure*}[t]
    \centering
    \begin{subfigure}[b]{0.32\textwidth}
        \centering
        \includegraphics[width=\linewidth]{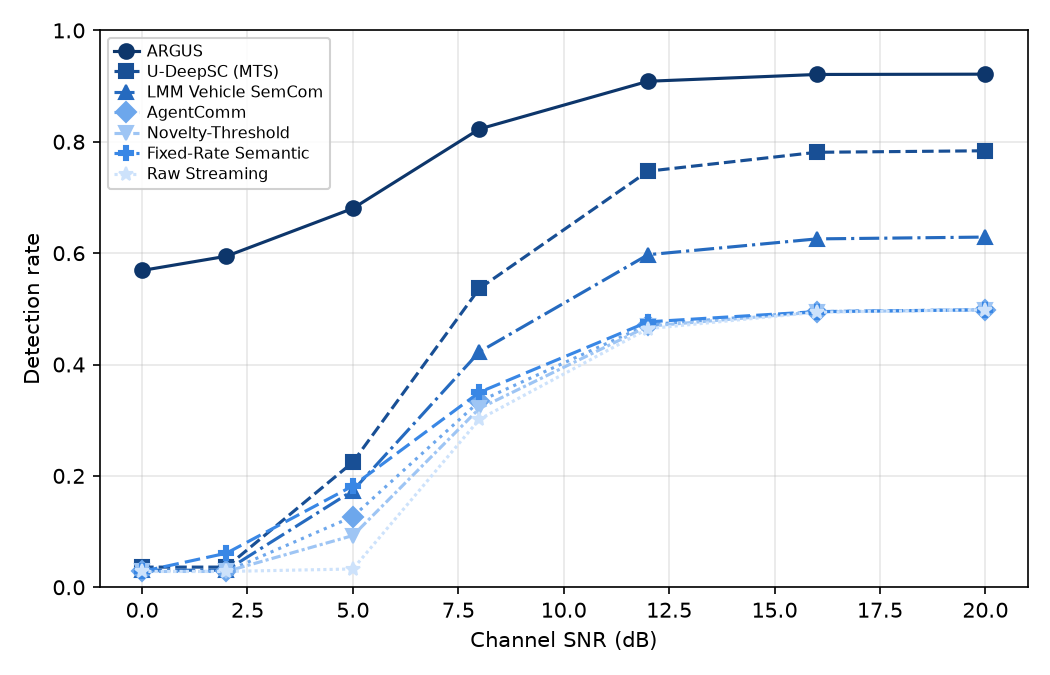}
        \caption{Detection rate vs. SNR.}
        \label{fig:snr_robustness}
    \end{subfigure}
    \hfill
    \begin{subfigure}[b]{0.32\textwidth}
        \centering
        \includegraphics[width=\linewidth]{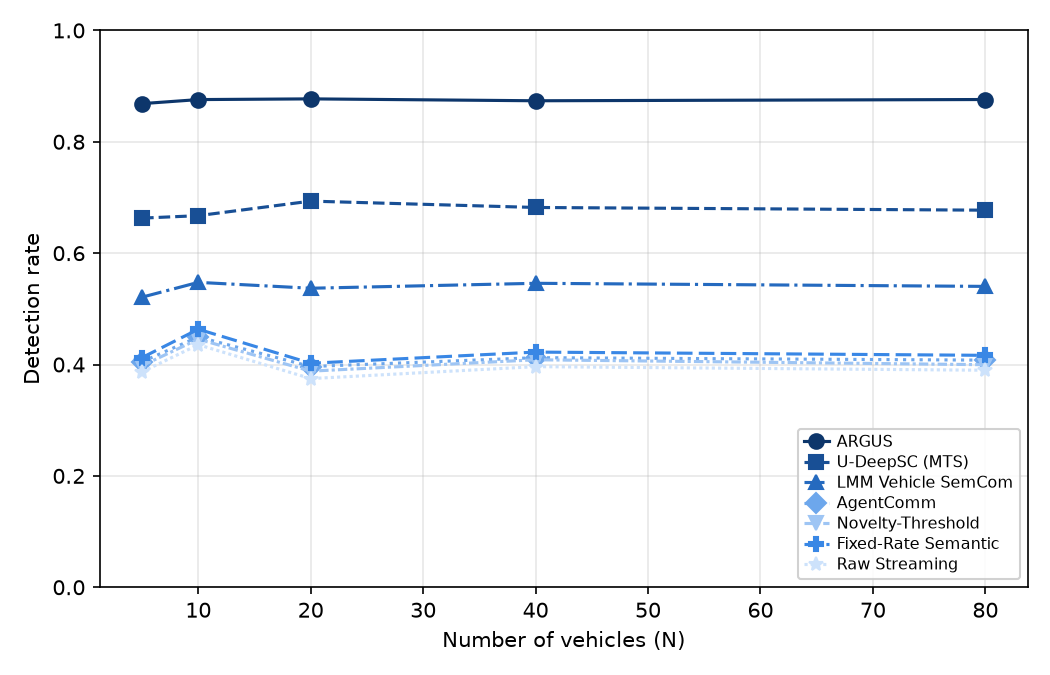}
        \caption{Detection rate vs. fleet size.}
        \label{fig:detvsfleet}
    \end{subfigure}
    \hfill
    \begin{subfigure}[b]{0.32\textwidth}
        \centering
        \includegraphics[width=\linewidth]{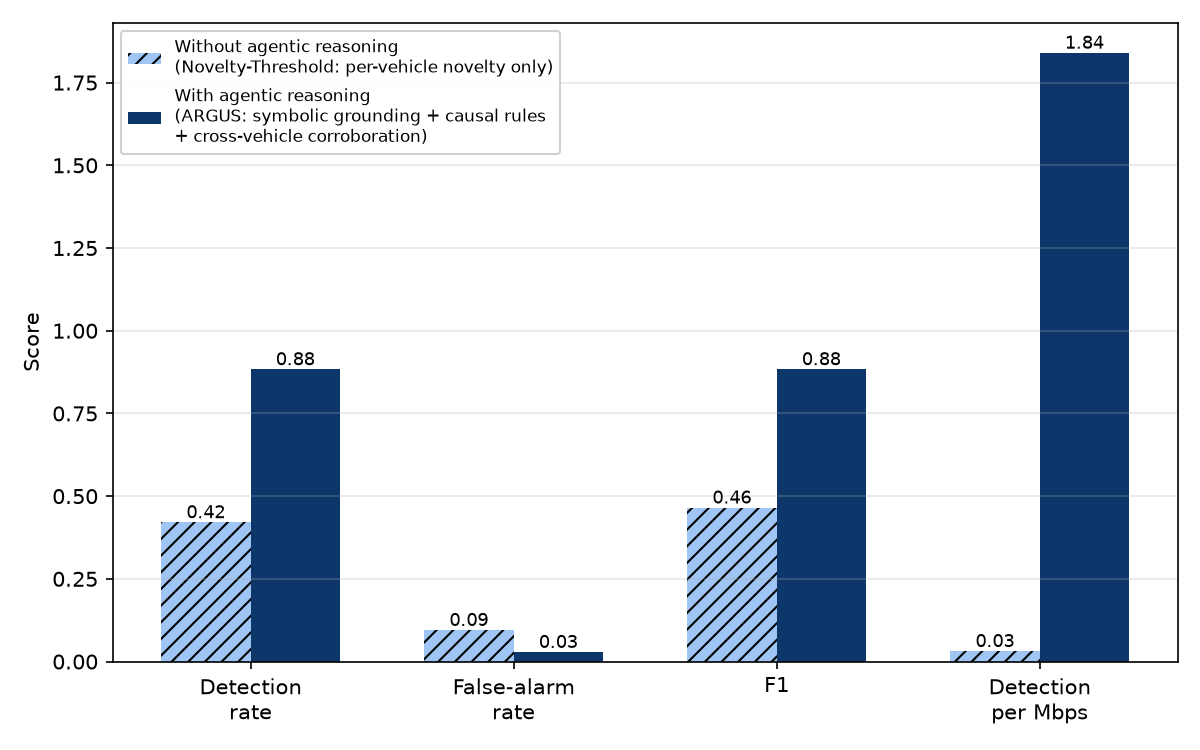}
        \caption{Agentic AI performance effect.}
        \label{fig:agentic}
    \end{subfigure}
    \caption{Experimental evaluation: (a), (b) Robustness and fleet scalability; and (c) Agentic AI effect of ARGUS.}
    \label{fig:combined_evaluation}
            \vspace{-0.4cm}
\end{figure*}

The performance of vehicular semantic communication depends on wireless channel conditions and fleet density. Evaluating robustness against channel fading and scalability across varying fleet sizes is therefore essential. Fig. ~\ref{fig:snr_robustness} examines detection performance under channel conditions ranging from 0 to 20~dB SNR. ARGUS maintains usable detection performance even at 0~dB and exceeds 90\% detection at approximately 12~dB, after which the performance remains relatively stable. U-DeepSC and LMM Vehicle SemCom approach approximately 78\% and 63\% detection, respectively, at favorable SNRs, while the remaining baselines saturate near 50\%. The improved robustness of ARGUS is consistent with the use of diffusion-based JSCC \cite{bourtsoulatze2019deep} for semantic transmission, which is designed to preserve task-relevant representations under channel impairments. The RSU-side collaborative reasoning further allows information from multiple observations to be integrated after reception, reducing the dependence on a single potentially corrupted transmission.

Fig.~\ref{fig:detvsfleet} shows the detection rate as the fleet size increases from 5 to 80 vehicles. The detection rate of ARGUS remains approximately stable at 0.87 across fleet sizes from 5 to 80 vehicles. The relatively stable detection rate indicates that the task-level performance is not substantially affected by the increase in fleet size over the evaluated range. At the same time, the approximately linear bandwidth growth observed in Fig.~\ref{fig:scaling} shows that this stability is achieved with an increasing aggregate communication load as the fleet expands.

\subsection{Latency and Transmission Reliability}

\begin{figure*}
\centering
\begin{subfigure}{0.33\textwidth}
    \includegraphics[width=\textwidth]{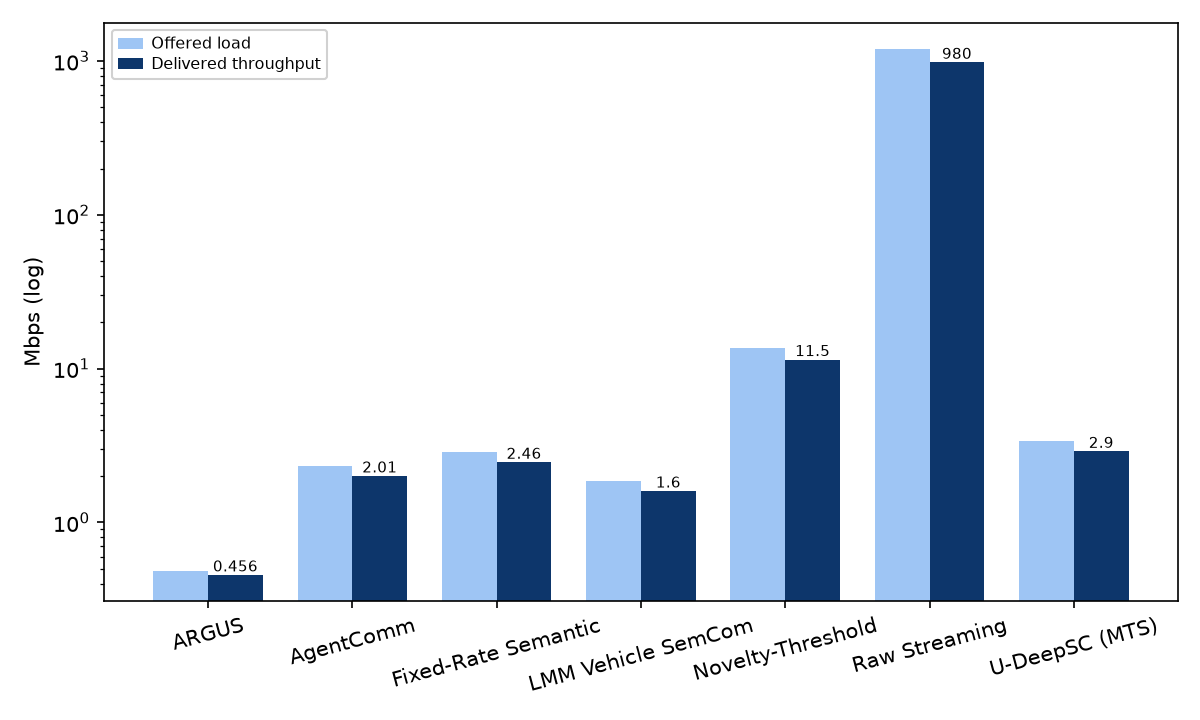}
    \caption{Offered load vs. delivered throughput.}
    \label{fig:throughput}
\end{subfigure}
\hfill
\begin{subfigure}{0.32\textwidth}
    \includegraphics[width=\textwidth]{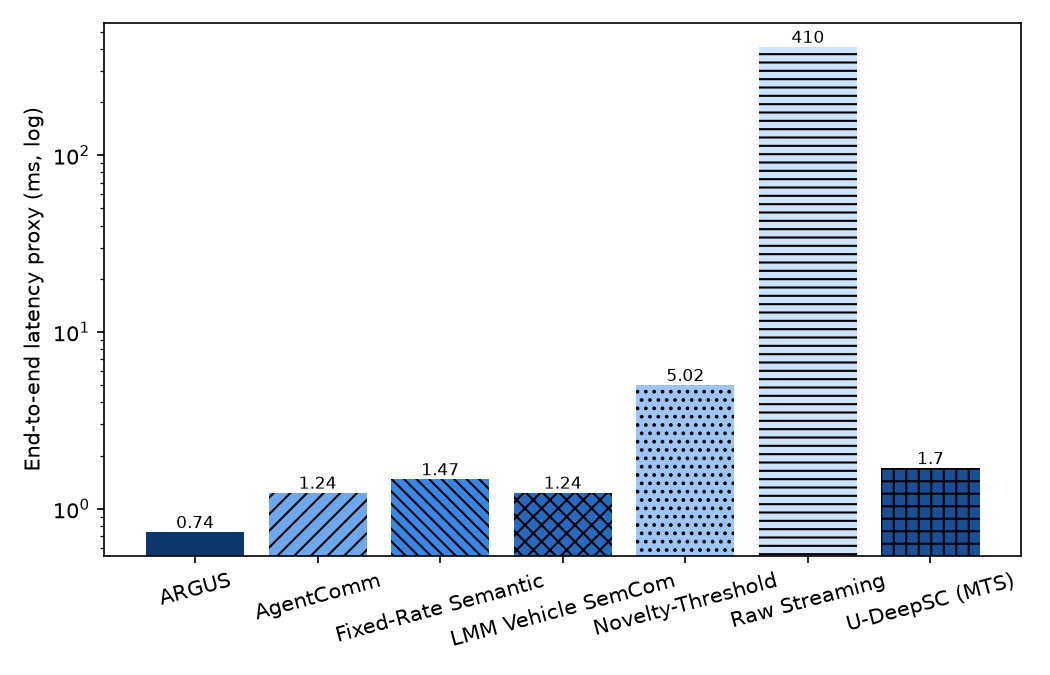}
    \caption{E2E latency proxy.}
    \label{fig:latency}
\end{subfigure}
\hfill
\begin{subfigure}{0.33\textwidth}
    \includegraphics[width=\textwidth]{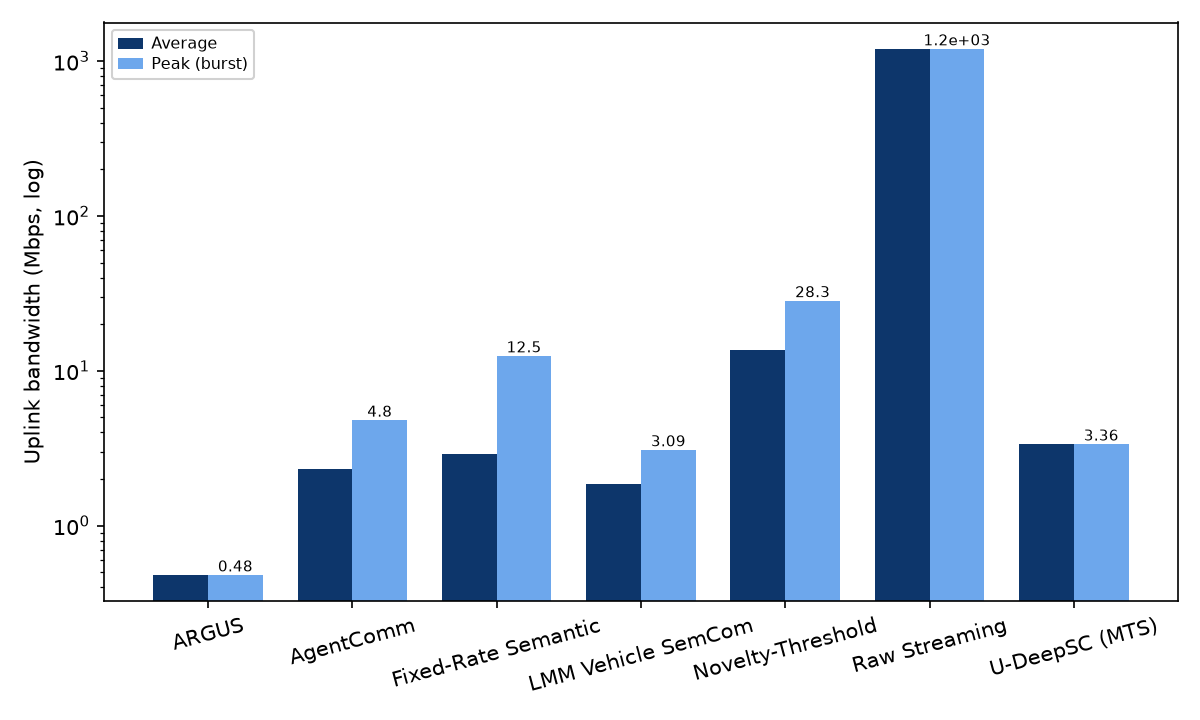}
    \caption{Average vs. peak uplink bandwidth.}
    \label{fig:burst}
\end{subfigure}
\caption{System-level throughput, latency and uplink bandwidth metrics.}
\label{fig:throughput_latency_burst}
        \vspace{-0.3cm}
\end{figure*}

The efficiency of information exchange is also reflected in the timeliness and reliability of semantic updates. Goodput characterizes the successfully delivered portion of the offered traffic, while latency and bandwidth burstiness capture the impact of transmission volume and temporal traffic variation on network operation. Fig.~\ref{fig:throughput} evaluates the fraction of offered communication that is successfully delivered. ARGUS converts approximately 0.48~Mbps of offered traffic into 0.456~Mbps of delivered goodput, corresponding to a 95.0\% delivery ratio. The other methods achieve approximately 82--87\%, while Raw Sensor Streaming reaches approximately 81.7\%. The higher effective delivery ratio observed for ARGUS is associated with its smaller and more regular semantic payloads, which place substantially less demand on the shared uplink than continuous high-volume transmission.

Fig.~\ref{fig:latency} compares the latency proxy associated with each communication strategy. ARGUS achieves approximately 0.74~ms, while the semantic baselines remain within the millisecond range. In contrast, Novelty-Thresholded SemCom increases to approximately 5.02~ms because novelty-triggered image patches introduce larger instantaneous payloads. Raw Sensor Streaming exhibits a much larger latency proxy of approximately 410~ms, reflecting the substantially greater volume of data that must be transmitted and processed. These results are consistent with the bandwidth measurements: reducing the amount of information transmitted per decision interval also reduces the communication burden associated with each semantic update.

Fig.~\ref{fig:burst} further illustrates this effect by comparing average and peak bandwidth. ARGUS maintains identical average and peak bandwidth at 0.48~Mbps because its regular semantic transmission policy limits instantaneous traffic variation. U-DeepSC exhibits similar behavior because it also uses a fixed-size representation. In contrast, adaptive event-driven approaches can generate substantially larger instantaneous loads when additional information is triggered; for example, Novelty-Thresholded SemCom increases from approximately 13.5~Mbps average to 28.3~Mbps peak. This difference is important in a shared C-V2X uplink, where bursty transmissions can temporarily increase contention even when their long-term average bandwidth is moderate.

\subsection{Computational Complexity}

\begin{figure*}[t]
    \centering
    \begin{subfigure}[b]{0.60\textwidth}
        \centering
        \includegraphics[width=\textwidth]{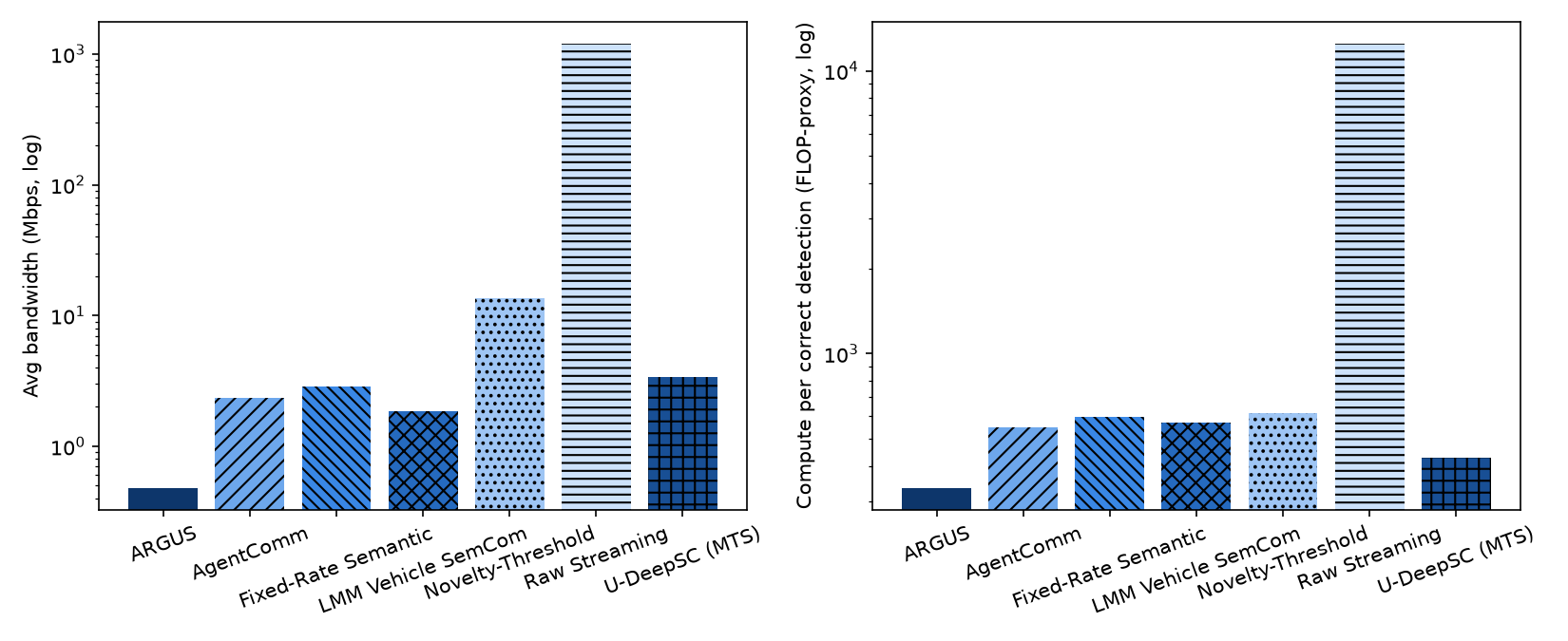}
        \caption{Communication and computational complexity.}
        \label{fig:complexity}
    \end{subfigure}
    \hfill
    \begin{subfigure}[b]{0.39\textwidth}
        \centering
        \includegraphics[width=\textwidth]{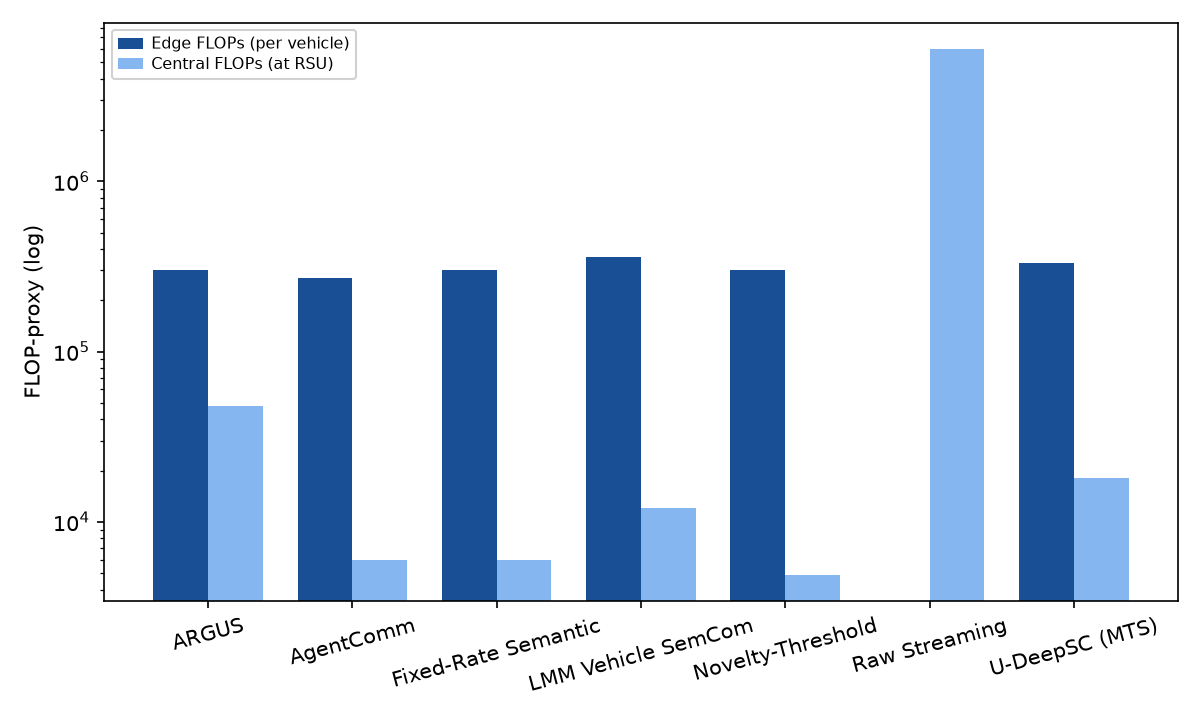}
        \caption{Edge vs. central FLOP-proxy per method.}
        \label{fig:compute}
    \end{subfigure}
    \caption{Computational cost and its distribution between vehicle-side and RSU-side processing.}
    \label{fig:complexity_compute}
            \vspace{-0.45cm}
\end{figure*}

The computational requirements of the considered approaches arise from different distributions of processing between the vehicle and the RSU. Separating vehicle-side and RSU-side computation provides insight into the computational implications of replacing raw-data transmission with semantic processing and network-level reasoning. Fig.~\ref{fig:complexity} evaluates computational cost relative to successful hazard detection. ARGUS requires approximately 335 floating-point operations (FLOPs) per correct detection, compared with 428 FLOPs/detection for U-DeepSC and higher values for the remaining semantic baselines. This metric reflects the combined effect of communication and processing costs relative to the number of correctly detected hazards; therefore, the lower value obtained by ARGUS is primarily associated with its substantially higher detection rate and reduced communication load. This does not imply that ARGUS has the lowest absolute computational cost. In fact, its centralized reasoning module requires more computation than several lightweight semantic baselines, as examined separately in Fig.~\ref{fig:compute}. Raw Sensor Streaming incurs the largest overall cost because the RSU processes high-volume video data rather than compact semantic representations.

Fig.~\ref{fig:compute} separates vehicle-side and RSU-side computational costs. For the semantic approaches, vehicle-side computation is dominated by VLM inference, whereas Raw Sensor Streaming avoids this semantic processing at the vehicle but shifts a much larger processing burden to the RSU. ARGUS requires approximately 48,000 FLOPs at the central node, compared with 18,000 FLOPs for U-DeepSC. Although ARGUS therefore incurs additional RSU-side reasoning cost, this cost remains far below the approximately 6,000,000 FLOPs required by Raw Sensor Streaming for full-frame processing. The result illustrates the intended design trade-off of ARGUS: additional edge-side semantic reasoning is used to reduce communication volume and improve task-level hazard detection. 

\subsection{Agentic AI effect}

Fig.~\ref{fig:agentic} isolates the contribution of ARGUS's agentic reasoning layer by comparing it against Novelty-Threshold, a baseline that shares the same adaptive, novelty-driven bandwidth philosophy but makes its hazard decision from a single per-vehicle novelty score alone. The gap between the two methods is attributable entirely to the reasoning layer ARGUS adds on top of that shared communication strategy: rather than treating "this scene looks unusual" as a proxy for "this scene is hazardous," ARGUS grounds each vehicle's caption into an explicit set of symbolic driving concepts and evaluates them against a causal rule base encoding domain knowledge about what combinations of entities constitute a hazard, then fuses corroborating reports from spatially neighboring vehicles before committing to a decision. This lets ARGUS decouple what is transmitted from what is dangerous, a distinction that a single novelty threshold cannot make.

This mechanism explains each of the four measured gains. Detection rate rises from 0.42 to 0.88 because causal rules fire on symbolic evidence (e.g., a pedestrian near an intersection, a stopped vehicle ahead) independently of whether the scene is visually novel, so ARGUS catches subtle hazards, real dangers that do not register as unexpected relative to a vehicle's recent history, that a novelty threshold misses entirely. FA rate falls from 0.095 to 0.028 because the converse also holds: a scene can be highly novel yet contain no grounded hazard symbol, and ARGUS does not flag it, whereas Novelty-Threshold flags any sufficiently unusual scene regardless of whether it is actually dangerous. Because detection and FA rate improve simultaneously rather than trading off against each other, F1 very nearly triples (0.465 to 0.883), reflecting a genuine joint improvement in both precision and recall rather than a shift along a single operating point. Finally, detection delivered per Mbps increases by roughly $60\times$ (0.03 to 1.84): ARGUS transmits small, fixed-size compact semantic message regardless of hazard severity, so none of its bandwidth is spent on large escalation payloads the way Novelty-Threshold's burst transmissions are, and the higher detection rate is achieved on top of that already-lower bandwidth footprint rather than at the cost of it. Taken together, these results confirm that ARGUS's gains originate from its agentic reasoning layer, symbolic grounding, causal rule evaluation, and cross-vehicle corroboration, and not merely from the adaptive bandwidth strategy it shares with the reactive baseline.

\section{Conclusion}

In this paper, we proposed ARGUS, an agentic semantic communication framework for adaptive and goal-driven information exchange in cooperative autonomous vehicle networks. 
ARGUS provides a unified system architecture that connects vehicle-side semantic perception, novelty-aware information selection, wireless transmission, and RSU-side reasoning and cross-vehicle information fusion. Rather than treating semantic communication as a mechanism for reducing the size of transmitted observations, ARGUS grounds semantic descriptions in structured driving concepts and uses this grounded representation to identify task-relevant and potentially safety-critical information. The resulting agentic architecture enables communication decisions to be guided by the evolving information needs of cooperative driving tasks, allowing redundant observations to be suppressed while prioritizing meaningful scene changes. The simulation results show that ARGUS reduces uplink bandwidth while maintaining high hazard-detection performance, low latency, and high communication efficiency compared to the baseline approaches evaluated.

\section*{Acknowledgment}

This work has been supported by MITACS, Ericsson Canada, and partially by NSERC Collaborative Research and Training Experience Program (CREATE) TRAVERSAL under Grant 497981 and Canada Research Chairs program. We acknowledge the use of generative AI tools to assist in the preparation of this manuscript. Specifically, OpenAI's GPT-6 Sol model and Anthropic's Claude Opus 4.8 model were used during the writing and simulation phase. All AI-generated text and code outputs were critically reviewed, modified, and validated by the human authors, who accept full responsibility for the overall accuracy and integrity of this work.

\bibliographystyle{IEEEtran}
\bibliography{ref}

\end{document}